%% file: main.tex
\documentclass[aip,amsmath,amssymb,reprint,
superscriptaddress
]{revtex4-2}

\usepackage{graphicx} 
\usepackage{hyperref}  
\usepackage[version=4]{mhchem} 
\usepackage{textcomp}
\usepackage{xcolor}
\usepackage{amsmath}
\usepackage{xspace}
\usepackage{cprotect}
\usepackage{comment}
\usepackage[T1]{fontenc}

\begin{document}

\include{macros}
\include{affiliations}


\title{
Abinit 2025: New Capabilities for the Predictive Modeling of Solids and Nanomaterials
}
\input{authors}
\date{June 2025}
\begin{abstract}
\abinit is a widely used scientific software package implementing density functional theory and many related functionalities for excited states and response properties. This paper presents the novel features and capabilities, both technical and scientific, which have been implemented over the past 5 years. This evolution occurred in the context of evolving hardware platforms, high-throughput calculation campaigns, and the growing use of machine learning to predict properties based on databases of first principles results. We present new methodologies for ground states with constrained charge, spin or temperature; 
for density functional perturbation theory extensions to flexoelectricity and polarons;
and for excited states in many-body frameworks including GW, dynamical mean field theory, and coupled cluster.
Technical advances have extended \abinit high-performance execution to graphical processing units and intensive parallelism.
Second principles methods build effective models on top of first principles results to scale up in length and time scales. Finally, workflows have been developed in different community frameworks to automate \abinit calculations and enable users to simulate hundreds or thousands of materials in controlled and reproducible conditions.
\end{abstract}

\maketitle

\input{1_introduction}
\input{2_groundstate}
\input{3_DFPT_EPC}
\input{4_excitationsandcorrelations}
\input{5_HPC}
\input{6_multiscale}

\input{7_highthroughput}
\input{8_conclusion}

\input{acknowledgements}
\bibliography{abinit2025}

\end{document}

%% file: macros.tex
\newcommand{\rtwoscan}{r$^2$SCAN}
\newcommand{\qq}{\mathbf{q}}
\newcommand{\enk}{\varepsilon_{n\kb}}
\newcommand{\emkq}{\varepsilon_{m\kb+\qb}}
\newcommand{\ef}{\varepsilon_F}
\newcommand{\qnu}{{{\qb\nu}}}
\newcommand{\wqnu}{{\omega_{\qnu}}}
\newcommand{\eqnu}{{\varepsilon_\qnu}}
\newcommand{\vnk}{\mathbf{v}_{n\kb}}
\newcommand{\vmkq}{\mathbf{v}_{m\kb+\qq}}
\newcommand{\vnka}{\mathrm{v}_{n\kb,\alpha}}
\newcommand{\vnkb}{\mathrm{v}_{n\kb,\beta}}
\newcommand{\gkq}{g_{mn\nu}(\kb,\qb)}
\newcommand{\mrtakkp}{\alpha_{mn\kb,\qb}}
\newcommand{\grid}[1]{${#1}\times{#1}\times{#1}$}
\newcommand{\ba}{{\mathbf a}}
\newcommand{\RR}{{\mathbf R}}
\newcommand{\rr}{{\mathbf r}}
\newcommand{\ie}{{\emph{i.e.}}\;}
\newcommand{\eg}{{\emph{e.g.}}\,\;}
\newcommand{\PDER}[2]{\dfrac{\partial #1}{\partial #2}}
\newcommand{\GG}{{\bf G}}
\newcommand{\bg}{{\bf g}}
\newcommand{\qG}{{\bf q+G}}
\newcommand{\kG}{{\bf k+G}}
\newcommand{\kg}{{\bf k+g}}
\newcommand{\kgp}{{\bf k+g'}}
\newcommand{\kpgp}{{\bf k'+g'}}
\newcommand{\FM}{{\text{FM}}}
\newcommand{\DW}{{\text{DW}}}
\newcommand{\KS}{{\text{KS}}}
\newcommand{\VH}{{V^{\text{H}}}}
\newcommand{\vH}{{v^{\text{H}}}}
\newcommand{\Vxc}{{V^{\text{xc}}}}
\newcommand{\vxc}{{v^{\text{xc}}}}
\newcommand{\Vscf}{{V^{\text{scf}}}}
\newcommand{\vscf}{{v^{\text{scf}}}}
\newcommand{\Vloc}{{V^{\text{loc}}}}
\newcommand{\vloc}{{v^{\text{loc}}}}
\newcommand{\Vnl}{{V^{\text{nl}}}}
\newcommand{\vnl}{{v^{\text{nl}}}}
\newcommand{\vks}{{v^\KS}}
\newcommand{\kk}{\mathbf{k}}
\newcommand{\kq}{\mathbf{k + q}}
\newcommand{\jj}{\mathbf{j}}
\newcommand{\vv}{\mathbf{v}}
\newcommand{\EE}{\mathbf{E}}
\newcommand{\bmcE}{{\mathbf{\mcE}}}
\newcommand{\bmcF}{{\mathbf{\mathcal{F}}}}
\newcommand{\FF}{\mathbf{F}}
\newcommand{\BB}{\mathbf{B}}
\newcommand{\TT}{\mathbf{T}}
\newcommand{\ww}{\omega}
\newcommand{\ab}{{\alpha\beta}}
\newcommand{\bme}{{\bm{\epsilon}}}
\newcommand{\dfodenk}{{\PDER{f^0}}}
\newcommand{\nk}{{n\kk}}
\newcommand{\mk}{{m\kk}}
\newcommand{\mkq}{{m\kq}}
\newcommand{\dd}{{\,\text{d}}}
\newcommand{\ket}[1]{|#1\rangle}
\newcommand{\bra}[1]{\langle#1|}
\newcommand{\braket}[2]{\langle#1 | #2 \rangle}
\newcommand{\abivar}[1]{\texttt{#1}}
\newcommand{\abinit}{\textsc{Abinit}\xspace}
\newcommand{\aiida}{AiiDA\xspace}
\newcommand{\mlacs}{\textsc{Mlacs}\xspace}
\newcommand{\atdep}{a\textsc{TDEP}\xspace}
\newcommand{\abioutfile}[1]{\textbf{#1}}
\newcommand{\abiexec}[1]{\texttt{#1}}
\newcommand{\abipy}{\textsc{AbiPy}\xspace}
\newcommand{\multibinit}{\textsc{Multibinit}\xspace}
\newcommand{\mcL}{\mathcal{L}}
\newcommand{\ee}{\varepsilon}

%% file: affiliations.tex
\newcommand{\ceadam}{\affiliation{CEA DAM-DIF, F-91297 Arpajon, France}}
\newcommand{\lmce}{\affiliation{Université Paris-Saclay, CEA, LMCE, F-91680 Bruyères-le-Châtel, France}}
\newcommand{\ceacadarache}{\affiliation{CEA, DES, IRESNE,DEC, Cadarache F 13108 St Paul Lez Durance, France}}
\newcommand{\ceaden}{\affiliation{Université Paris-Saclay, CEA, Service de recherche en Corrosion et Comportement des Matériaux, SRMP, 91191 Gif-sur-Yvette, France}}
\newcommand{\chalmers}{\affiliation{Dept. of Physics, Chalmers University of Technology, SE-412 96 Gothenburg, Sweden}}
\newcommand{\corning}{\affiliation{Corning Incorporated, SP-FR-05, Sullivan Park, Corning, NY 14831, USA}}
\newcommand{\dalhousie}{\affiliation{Dept. of Chemistry, Dalhousie University, Halifax, NS, B3H 4R2 Canada}}
\newcommand{\etsf}{\affiliation{European Theoretical Spectroscopy Facility, www.etsf.eu}}
\newcommand{\icmab}{\affiliation{Institut de Ci\`encia de Materials de Barcelona (ICMAB-CSIC), Campus UAB, 08193 Bellaterra, Spain}}
\newcommand{\icn}{\affiliation{Catalan Institute of Nanoscience and Nanotechnology (ICN2), Campus UAB, Bellaterra, 08193 Barcelona, Spain}}
\newcommand{\icrea}{\affiliation{ICREA - Instituci\'o Catalana de Recerca i Estudis Avan\c{c}ats, 08010 Barcelona, Spain}}
\newcommand{\imec}{\affiliation{Imec, Kapeldreef 75, B-3001 Leuven, Belgium}}
\newcommand{\impmc}{\affiliation{IMPMC, UMR 7590 of Sorbonne Universit\'e/CNRS/MNHN/IRD, Paris, France}}
\newcommand{\lsim}{\affiliation{IRIG-MEM, L-SIM , University Grenoble Alpes, CEA , F-38000 Grenoble , France}}
\newcommand{\matgenix}{\affiliation{Matgenix, A6K Advanced Engineering Centre, B-6000 Charleroi, Belgium}}
\newcommand{\matsim}{\affiliation{Mat-Sim Research LLC, Lafayette, CO 80026, USA}}
\newcommand{\montreal}{\affiliation{Dept. de Physique, U. de Montr\'{e}al, C.P. 6128, Succursale Centre-Ville, Montr\'{e}al H3C 3J7, Canada}}
\newcommand{\nanomat}{\affiliation{NanoMat/Q-Mat, Universit{\'e} de Li{\`e}ge (B5), B-4000 Li{\`e}ge, Belgium}}
\newcommand{\phythema}{\affiliation{Theoretical Materials Physics, Q-MAT, Universit{\'e} de Li{\`e}ge (B5a), B-4000 Sart-Tilman, Belgium}}
\newcommand{\rpi}{\affiliation{Dept. of Physics, Applied Physics, and Astronomy, Rensselaer Polytechnic Institute, Troy, New York 12180, USA}}
\newcommand{\rutgers}{\affiliation{Dept. of Physics and Astronomy, Rutgers U., Piscataway, NJ 08854-8019, USA}}
\newcommand{\uclouvain}{\affiliation{Institute of Condensed Matter and Nanoscience, UCLouvain, B-1348 Louvain-la-Neuve, Belgium}}
\newcommand{\uqtr}{\affiliation{Université du Qu\'ebec \`a Trois-Rivi\`eres, Institut de recherche sur l'hydrog\`ene, Trois-Rivi\`eres, Canada}}
\newcommand{\utrecht}{\affiliation{ITP, Dept of Physics, University of Utrecht, 3508 TA Utrecht, The Netherlands}}
\newcommand{\welt}{\affiliation{WEL Research Institute, avenue Pasteur 6, 1300 Wavre, Belgium}}
\newcommand{\wvirginia}{\affiliation{Physics and Astronomy Dept, West Virginia University, Morgantown, USA}}
\newcommand{\lcpq}{\affiliation{Laboratoire de Chimie et Physique Quantiques (UMR 5626), Universit\'e de Toulouse, CNRS, UPS, France}}
\newcommand{\uark}{\affiliation{Smart Ferroic Materials center, Institute of Nanoscience \& Engineering and Department of Physics, University of Arkansas, Fayetteville AR 72701, USA}}
\newcommand{\tauisr}{\affiliation{Department of Materials Science and Engineering, Tel Aviv University, Ramat Aviv, Tel Aviv 6997801, Israel}}
\newcommand{\spms}{\affiliation{Université Paris-Saclay, CentraleSupélec, CNRS, Laboratoire SPMS, 91190, Gif-sur-Yvette, France}}
\newcommand{\ccq}{\affiliation{Center for Computational Quantum Physics, Flatiron Institute, 162 Fifth Avenue, New York, New York 10010, USA}}
\newcommand{\ipht}{\affiliation{Université Paris-Saclay, CNRS, CEA, Institut de physique théorique, 91191, Gif-sur-Yvette, France}}
\newcommand{\kfupm}{\affiliation{Center for Integrative Petroleum Research, King Fahd University of Petroleum and Minerals, Dhahran 31261, Saudi Arabia}}
\newcommand{\wut}{\affiliation{School of Physics and Mechanics, Wuhan University of Technology, Wuhan 430070, China}}
\newcommand{\euxfel}{\affiliation{European XFEL, Holzkoppel 4, 22869 Schenefeld, Germany}}
\newcommand{\trinity}{\affiliation{School of Physics, Trinity College Dublin, The University of Dublin, Ireland}}
\newcommand{\asplus}{\affiliation{Alliance Service Plus. 62 rue Emile Zola. 91200 Boulogne-Billancourt. France}}
\newcommand{\wfu}{\affiliation{Department of Physics, Wake Forest University, Winston-Salem NC 27109 USA}}
\newcommand{\epflimx}{\affiliation{Institute of Materials Science and Engineering, École Polytechnique Fédérale de Lausanne, Station 1, 1015 Lausanne, Switzerland}}
\newcommand{\weizmann}{\affiliation{Dept of Molecular Chemistry and Materials Science, Weizmann Institute of Science, Rehovoth 76100, Israel}}
\newcommand{\tuwien}{\affiliation{
Institute for Theoretical Physics, TU Wien,
Wiedner Hauptstra{\ss}e 8--10/136, 1040 Vienna, Austria}}

%% file: authors.tex
\author{Matthieu J. Verstraete} 
\nanomat \utrecht \etsf
\author{Joao Abreu} 
\nanomat
\author{Guillaume E. Allemand} 
\nanomat \etsf
\author{Bernard Amadon} 
\ceadam \lmce
\author{Gabriel Antonius} 
\uqtr
\author{Maryam Azizi} 
\uclouvain \etsf
\author{Lucas Baguet} 
\ceadam \lmce
\author{Cl\'ementine Barat} 
\ceadam \lmce
\author{Louis Bastogne} 
\phythema
\author{Romuald B\'ejaud} 
\ceadam \lmce
\author{Jean-Michel Beuken} 
\uclouvain
\author{Jordan Bieder} 
\phythema
\author{Augustin Blanchet} 
\ceadam \lmce
\author{Francois Bottin} 
\ceadam \lmce
\author{Johann Bouchet} 
\ceacadarache
\author{Julien Bouquiaux} 
\uclouvain \matgenix
\author{Eric Bousquet} 
\phythema
\author{James Boust} 
\ceadam \lmce
\author{Fabien Brieuc} 
\ceadam \lmce
\author{V\'eronique Brousseau-Couture} 
\montreal
\author{Nils Brouwer} 
\euxfel
\author{Fabien Bruneval} 
\ceaden
\author{Alo\"is Castellano} 
\nanomat \etsf
\author{Emmanuel Castiel} 
\ceadam \lmce
\author{Jean-Baptiste Charraud} 
\ceadam \lmce
\author{Jean Cl\'erouin} 
\ceadam \lmce 
\author{Michel C\^{o}t\'{e}} 
\montreal
\author{Cl\'ement Duval} 
\ceadam \lmce
\author{Alejandro Gallo} 
\tuwien
\author{Frederic Gendron} 
\ceadam \lmce
\author{Gregory Geneste} 
\ceadam \lmce
\author{Philippe Ghosez} 
\phythema
\author{Matteo Giantomassi} 
\uclouvain \etsf
\author{Olivier Gingras} 
\ccq \ipht
\author{Fernando G\'omez-Ortiz} 
\phythema
\author{Xavier Gonze} 
\uclouvain \etsf
\author{F\'{e}lix Antoine Goudreault} 
\montreal
\author{Andreas Gr\"uneis} 
\tuwien
\author{Raveena Gupta} 
\nanomat
\author{Bogdan Guster} 
\phythema
\author{Donald R. Hamann} 
\rutgers \matsim

\author{Xu He} 
 \phythema

\author{Olle Hellman} 
\weizmann
\author{Natalie Holzwarth} 
\wfu
\author{Francois Jollet} 
\ceadam \lmce
\author{Pierre Kestener} 
\ceadam
\author{Ioanna-Maria Lygatsika} 
\ceadam \lmce
\author{Olivier Nadeau} 
\ceadam \lmce \uqtr
\author{Lórien MacEnulty} 
\trinity
\author{Enrico Marazzi} 
\uclouvain
\author{Maxime Mignolet} 
\nanomat \etsf
\author{David D. O'Regan} 
\trinity
\author{Robinson Outerovitch} 
\ceadam
\author{Charles Paillard} 
\uark \spms
\author{Guido Petretto} 
\matgenix\uclouvain
\author{Samuel Ponc\'e} 
\uclouvain \welt \etsf
\author{Francesco Ricci} 
\uclouvain \matgenix
\author{Gian-Marco Rignanese} 
\uclouvain \welt \etsf
\author{Mauricio Rodriguez-Mayorga} 
\lcpq
\author{Aldo H. Romero} 
\wvirginia
\author{Samare Rostami} 
\uclouvain \etsf
\author{Miquel Royo} 
\icmab
\author{Marc Sarraute} 
\asplus
\author{Alireza Sasani} 
\phythema

\author{Fran\c{c}ois Soubiran} 
\ceadam \lmce
\author{Massimiliano Stengel}
\icmab
\author{Christian Tantardini} 
\kfupm
\author{Marc Torrent} 
\ceadam \lmce
\author{Victor Trinquet} 
\uclouvain
\author{Vasilii Vasilchencko} 
\uclouvain
\author{David Waroquiers} 
\uclouvain \matgenix
\author{Asier Zabalo} 
\icmab
\author{Austin Zadoks} 
\epflimx
\author{Huazhang Zhang} 
\phythema \wut
\author{Josef Zwanziger} 
\dalhousie

%% file: 1_introduction.tex
\section{Introduction}

\label{sect:general_intro}


Ab initio calculations are based on numerical methods developed in quantum chemistry and computational materials science to predict molecular and material properties using fundamental physical constants without relying on empirical data. 
Over the past two decades, they have become a reliable and systematic tool to complement experiments and understanding of the microscopic mechanisms behind material functionality.
The most common framework solves the electronic Kohn-Sham equations derived from density functional theory (DFT), giving access to many properties such as the electron wave functions, band structure, bonding characteristics, or chemical reactivity
~\cite{Martin2004, Kohanoff2006, Giustino2014}. 
As computational power has increased exponentially through the use of CPUs to GPUs, high parallelization, and better computer node communication, the development of electronic structure codes has become pivotal, enabling scientists to perform ever more complex simulations with greater accuracy, a more significant number of degrees of freedom, and efficiency~\cite{Gavini2023}. 
At the turn of the century, the field was democratized and amplified by the open-source release of several general-purpose software packages, starting with \abinit~\cite{Gonze2002}, Quantum ESPRESSO~\cite{Giannozzi2009}, SIESTA~\cite{Garcia2020}, then exciting~\cite{Gulans2014}/Elk~\cite{elk} and many others. 
These tools implement a variety of quantum mechanical models and algorithms, facilitating comprehensive studies across physics, chemistry, and materials science with robust capabilities in predicting electronic, elastic, and magnetic material properties. 
The most widely used method combines plane-wave basis sets, with pseudopotentials or the projector augmented wave (PAW) method, to solve the Kohn-Sham equations~\cite{Martin2004}.
Over the years, such calculations have been used to compute many properties of materials in a wide range of conditions and study various phenomena at the electronic and atomic scale. It has empowered scientists to investigate millions of materials, enabling machine learning algorithmic exploration of chemical space, and the efficient distillation of candidate materials before experimental investigation.

In this paper, we focus on the recent advances in the \abinit software suite, which has a lively user community and an engaged group of developers. 
The project began in 1997 with the vision of providing a free, collaborative platform for ab initio calculations. 
Its initial public release occurred in December 2000, marking the first open source availability of a tool that implements DFT~\cite{Gonze2002}. 
Over the years, \abinit has evolved through contributions from an international community of researchers
~\cite{Gonze2005,Gonze2009, Gonze2016, Gonze2020, Romero2020}
. Significant milestones include the implementation of density functional perturbation theory (DFPT) to calculate response functions such as phonons, electron-phonon coupling, Raman spectra, and dielectric responses~\cite{Baroni2001,Gonze2005a}, as well as the incorporation of advanced electronic excited state methods like 
the $GW$ approximation, the Bethe-Salpeter equation, and the dynamical mean-field theory (DMFT)~\cite{Kotliar2006,Martin2016,Golze2019}. 
These developments have expanded \abinit capabilities to encompass almost all known material properties, including structural, electronic, vibrational, and thermodynamic characteristics. Thanks to the project's collaborative nature, new developers can change the package and include features with very clear workflows, documentation, testing, and computational engineering design. \abinit development has fostered a robust and versatile tool which has been widely adopted in the scientific community for research, but also educational purposes, as it sports a very well-supported set of tutorials with theoretical and computational details.

In the last five years, since our publications in 2020~\cite{Romero2020, Gonze2020}, \abinit have undergone significant developments, enhancing its capabilities and performance across various aspects of electronic structure calculations. Recent improvements in ground-state computations have resulted in more accurate and efficient predictions of electronic properties in diverse materials and thermodynamic conditions. 
The present paper discusses our constrained density functional theory (cDFT) implementation, which allows explicit constraints on electronic densities and provides a refined description of excited states, charge transfers, and localized excitations.
We address challenges posed by high-temperature extreme conditions, where standard DFT implementations are inadequate, by integrating thermal exchange-correlation functionals. These new functionals extend DFT applicability to warm dense matter and plasmas, inviting broader interdisciplinary collaboration.
\abinit has also introduced methodologies for computing long-wave properties, including lattice-mediated bulk flexoelectricity and dielectric screening in quasi-two-dimensional crystals. These improvements enable studies of gradient-dependent effects like natural optical rotation and polarization responses to strain gradients, which are crucial for understanding phenomena such as flexoelectricity. 
Investigations into phonon angular momentum provide essential insights into the interactions between the lattice vibrations, light, magnetic fields, and electronic structure, which are particularly significant in chiral crystals~\cite{bousquet2024structural}. 
Incorporating the $GW$ one-body reduced density matrix offers a more accurate representation of electronic correlations, complemented by enhancements to our dynamical mean-field theory (DMFT) implementation. These updates bolster \abinit's capacity to study strongly correlated electron systems. 

On the more technical front, significant improvements have also been made in \abinit's user interface and installation processes, accommodating configurations into a diversity of large-scale supercomputers. The command-line interface (CLI) has been enhanced with numerous options, comprehensively listed and described through the command \abiexec{abinit --help}. Users can now access detailed information about the build environment, including utilized libraries, compilation options, and hardware specifications, and directly manage runtime parameters such as thread counts, execution limits, and parallelization modes. The functional design of input files has been substantially improved, notably by allowing parameters to be set using intuitive character strings instead of opaque integer codes. Output files have been enriched with structured sections written in yaml~\cite{yaml_website}, greatly facilitating their parsing by post-processing scripts, particularly those from the \abipy~\cite{abipy_website} python package. Clear labeling of primary results sections has enabled the implementation of automated non-regression and validation tests grouped under the \texttt{minimal} test suite, ensuring reliable validation independent of hardware architecture and compilation settings.
Regarding code compilation, the principal advancement is adopting a build system based on \texttt{cmake}~\cite{cmake_website}, in parallel with the established \texttt{autotools}-based reference system. The detailed use of this modern build system is documented comprehensively. For users who prefer to bypass direct code compilation, \abinit is available through package managers like \texttt{EasyBuild}~\cite{easybuild_website}, \texttt{Spack}~\cite{spack_website}, \texttt{Homebrew}~\cite{homebrew_website}, and \texttt{Anaconda}~\cite{anaconda}. These package managers either download pre-compiled executables or automate compilation to optimize performance while managing dependencies seamlessly.
Given the increasing complexity of host architectures involving diverse compilers, performance libraries, and graphical accelerators, compiling and installing \abinit demands technical sophistication. Performance across various architectures has significantly improved since 2020 (see section~\ref{sect:HPC}), emphasizing users' need to master the compilation chain to leverage these performance gains fully.

The field of electronic structure calculations is increasingly integrating high performance, high-throughput, and artificial intelligence methodologies.
Advances in high-performance computing are being leveraged to tackle ever more complex and precise simulations.
The increasing complexity of calculations and materials (with defects, under extreme conditions, etc.) has encouraged interest in developing machine learning models that can predict material properties efficiently and correlate them with crystal structure. 
This shift necessitates the generation of extensive, reliable datasets encompassing a wide array of observables, including total energies, convex hulls, structural stability, phonons, Raman and infrared spectra, and X-ray properties. We finish this article with a section on workflows and machinery for the automation of calculations. Note that the \abinit code uses Hartree atomic units ($m_e = e = \hbar = 4\pi\epsilon_0 = 1$, 1~Ha = 27.211~eV) internally for all expressions, and these will be the default below, unless otherwise stated.

%% file: 2_groundstate.tex
\section{Novel Features for Ground 
State Calculations}
\label{sect:GS}


Computing the ground state in Kohn-Sham DFT has been a feature of \abinit since
its inception. Nevertheless, as the range of applicability of DFT has increased
over time, new features and targeted behavior for the ground state calculation
were added to \abinit. Here we describe several capabilities that
have been implemented since the last overview papers~\cite{Gonze2020,Romero2020}.
These include the implementation of constraints on atomic charge and magnetization;
constraints on band filling, to model photodoping;
high-temperature DFT; 
availability of meta-GGA functionals in the projector augmented-wave and norm-conserving pseudopotential frameworks;
and orbital magnetism and magnetic shielding.


\subsection{Constrained DFT}
\label{subsect:GS_constraint}

Constrained density functional theory can be used to impose constraints on atomic charges and magnetization.
Usually, the integrals of the charge density and magnetization (real-space functions) inside predefined spheres are constrained to user-defined values. This is described in, {\em e.g.}, Refs.~\citenum{Kaduk2012,Ma2015}.

Two cDFT al\-go\-rithms have been implemented in \textsc{Abinit}.
The first al\-go\-rithm is the one presented in Ref.~\citenum{Ma2015}, 
based on a penalty function.
It imposes the target values approximately, under the control of \abivar{magcon\_lambda}, the penalty function prefactor.
This algorithm has been available already since before {\abinit}v9. 
In line with the description in Ref.~\citenum{Ma2015}, it is implemented in \abinit only for the atomic magnetic moments.

The second, more recent algorithm, is based on a Potential-based self-consistency loop and the Lagrange multiplier (LM) method, and is referred to as PLM-cDFT hereafter. It is described in Ref.~\citenum{Gonze2022}. PLM-cDFT is activated with the \abivar{constraint\_kind} input variable.
The constraints of five quantities are implemented: the atomic charge, the vector atomic magnetization, the atomic magnetization length, the atomic magnetization axis ({\em e.g.}, along $\pm z$) and the atomic magnetization direction ({\em e.g.}, along $+z$). 
Each type of atom can have these activated either separately or simultaneously.

PLM-cDFT is an improvement over both algorithms from Refs.~\citenum{Kaduk2012, Ma2015}.
Compared to the penalty function approach in Ref.~\citenum{Ma2015}, LMs impose the constraints to arbitrary numerical precision.
This proved essential in order to generate magnetic 
moment tensor potentials in recent 
publications~\cite{Kotykhov2023a,Kotykhov2024,Kotykhov2025}.

Moreover, PLM-cDFT has recently been demonstrated to resolve the long-standing discrepancy between theory and experiment for the ferrimagnetic kagome oxide \ce{YBaCo4O7} (Y114). 
Using potential-based self-consistent charge constraints on the Co sublattice and enforcing zero magnetization on the Y, Ba, and O ions, Ref.~\citenum{Tantardinin2025_Y114} reproduces the experimentally observed Co$^{2+}$/Co$^{3+}$ charge disproportionation and obtains Co magnetic moments that agree within 0.1 $\mu\mathrm{B}$ with neutron data, outperforming conventional DFT+$U(+J)$ approaches. 
This application highlights how specifying physically transparent oxidation-state constraints in PLM-cDFT eliminates the spurious magnetization leakage to nominally non-magnetic ions and yields a quantitatively reliable description of complex magnetic materials.

PLM-cDFT is also not subject to instabilities that have been observed when \abivar{magcon\_lambda} becomes large based
on the algorithm of Ref.~\citenum{Ma2015}. Still, for GGA functionals, the PLM-cDFT is not unconditionally stable, and this is currently under analysis.

In the PLM-cDFT algorithm, forces as well as stresses can be computed. Also, derivatives of the total energy with respect to the constraint are delivered.
For four types of constraints (atomic charge, vector atomic magnetization, the atomic magnetization length 
and the atomic magnetization axis), the derivative
is computed with respect to the value of the constraint defined directly by the user, while for the magnetization direction constraint,
the derivative is evaluated with respect to the change of angle.

If \abivar{constraint\_kind} is non-zero for at least one type of atom, the PLM-cDFT algorithm 
\abivar{constraint\_kind} defines, for each type of atom, the kind of constraint(s) imposed by cDFT. When \abivar{constraint\_kind} is zero for an atom type, there is no constraint applied to this atom type. Otherwise, different constraints can be imposed on the total charge (ion+electronic) and/or magnetization, computed inside a sphere of radius
set by \abivar{ratsph}, possibly smeared within a width set by \abivar{ratsm}. The integrated ion plus
electronic charge is constrained to be equal to 
the value set by \abivar{chrgat}, while the magnetization is constrained to be equal to (or collinear with) the vector
set by \abivar{spinat}.

\subsection{Quasi-Fermi energies and photodoping}
\label{subsect:GS_photodoping}
To mimic photo-induced structural deformations or even phase transitions, a constrained occupation number DFT scheme has been implemented in \abinit~\cite{Paillard2019} since version 9.4, which can be activated using \abivar{occopt} = 9. In this case, we separate the valence and conduction Kohn-Sham states by providing the valence band index \abivar{ivalence} in the input file, and define two quasi-Fermi energies $\varepsilon_e$ and $\varepsilon_h$ which control the filling of the Kohn-Sham occupation numbers in the conduction and valence bands, respectively. This method mimics the effect of thermalized photo-excited carriers.
Beyond \abivar{occopt 9} and the \abivar{ivalence} tag, the user must input the number $n_{ph}$ of photodoped electrons constrained in the conduction bands, using the tag \abivar{nph}. Note that the number of holes in the valence bands ({\em i.e.}, the lowest $N_v$ Kohn-Sham states identified by \abivar{ivalence}), is currently automatically set equal to the number of electrons in the conduction bands to ensure charge neutrality of the system.

At each self-consistent field iteration, the quasi-Fermi energies $\ee_e$ and $\ee_h$ are found using a bisection algorithm (as is done in the ground state for $3 \leq$ \abivar{occopt}  $\leq 8$) by solving the following equations
\begin{equation}
\label{eq:fermi_holes}
N-n_{ph}  = \sum_{n \leq N_v} \sum_{\kk, s} w_\kk \tilde{\theta}\left( \frac{\varepsilon_{n\kk s} - \varepsilon_h}{\sigma} \right),
\end{equation}
\begin{equation}
\label{eq:fermi_electrons}
n_{ph}    =  \sum_{n > N_v} \sum_{\kk, s} w_\kk \tilde{\theta}\left( \frac{\varepsilon_{n\kk s} - \varepsilon_e}{\sigma} \right).
\end{equation}
Note that in Eqs.~\ref{eq:fermi_holes}-\ref{eq:fermi_electrons}, $w_{\kk}$ is the weight of the $\kk$ wavevector in the irreducible Brillouin zone, $s$ represents the spin index, $\varepsilon_{n\kk s}$ is the Kohn-Sham eigenvalue of state $n$ at vector $\kk$ with spin $s$, and $N$ is the total number of electrons in the calculated cell. $\tilde{\theta}$ is the smearing function and $\sigma$ is the smearing temperature (defined by the user using the \abivar{tsmear} tag in the input file). Note that at present, only the Fermi-Dirac smearing scheme is available, but can be generalized to other metallic smearing schemes in the future.

Subsequently, the occupation numbers of valence (resp. conduction) band states are filled using the smearing scheme with their own quasi-Fermi level $\varepsilon_h$ (resp. $\varepsilon_e$), ensuring that $n_{ph}$ holes (resp. electrons) are constrained in the valence (resp. conduction) bands, and the output density of the self consistent field iteration is calculated with these occupation numbers.

At present, this scheme allows to perform structural relaxation in non-spin polarized, spin-polarized and spinorial calculations, within DFT or DFT$+U$ in \abinit.  The density functional perturbation theory version of this formalism has been recently elaborated by Marini and Calandra~\cite{Marini2021} and is currently being implemented in \abinit.


\subsection{High-temperature DFT}
\label{subsect:GS_highT}

\subsubsection{Extended DFT}

The Mermin finite-temperature formulation of Kohn-Sham DFT has proven to be
successful in handling the
complexity of the warm dense matter regime, for electronic temperatures
close to the Fermi temperature of the material
$T_\mathrm{F} = E_\mathrm{F}/k_\mathrm{B}$ and even above it.
Unfortunately, plane-wave-based DFT is rather limited to low
temperatures, because of the \emph{orbital wall}~\cite{Blanchet2020}: at high temperatures,
the Fermi-Dirac distribution imposes to consider a large number
of weakly occupied states in the dense high energy continuum.
Computing electronic properties, such as the equation of state,
for temperatures higher than $T_\mathrm{F}$ is then completely
unreachable, even on the largest supercomputers.

This issue is well known in the DFT plasma community.
Among the emerging models that claim to bypass the
\emph{orbital wall}~\cite{Baer2013, Suryanarayana2018, Chabrier2019, White2020, Militzer2021, Sharma2023, Bethkenhagen2023}, the extended
plane-waves DFT model, originally named ``extended first-principles MD''~\cite{Zhang2016, Blanchet2022, Hollebon2022} has been successful. It replaces high-energy electronic states with pure plane
waves, approximating that electrons occupying such high energy
states behave like free electrons. As an example, the total electron
density $n(\textbf{r})$ is re-expressed as:
\begin{equation}
  \label{eq:extfpmd_density}
  n(\mathbf r) = 2\sum_\mathbf{k} w_\mathbf{k}
  \sum_{n=1}^\text{\abivar{nband}} f_{n\mathbf{k}}
  \left|\psi_{n\mathbf{k}}(\mathbf r)\right|^2 +
  \frac{1}{\Omega} N^\mathrm{ext},
\end{equation}
with $f_{n\mathbf{k}} = f(\ee_{n\mathbf{k}}) =
(e^{\beta (\ee_{n\mathbf{k}} - \mu)} + 1)^{-1}$ the Fermi-Dirac function, $\mu$ the electronic chemical
potential, $w_\mathbf{k}$ the weight of the $\mathbf{k}$ wavevector, $\Omega$ the unit cell volume, $n$ the band index, $\mathbf{k}$ the wavevectors and $\ee_{n\mathbf{k}}$ the KS eigenenergies. The number of electrons described with the extended DFT model $N^\mathrm{ext}$ is expressed as
\begin{equation}
  N^\mathrm{ext} = \frac{\sqrt{2}}{\pi^2} \Omega
  \int_{\ee_\mathrm{c}}^\infty f(\epsilon)
  \sqrt{\ee - U_0} \, d\ee,
\end{equation}
where $\ee_\mathrm{c} = \sum_\mathbf{k} w_\mathbf{k}
\epsilon_{\text{\abivar{nband}}\mathbf{k}}$, $U_0 = \int v(\mathbf{r}) \, d^3\mathbf{r}
/ \Omega$ and $v(\mathbf{r})$ is the local part of the KS potential.

The contributions of the extended DFT model to the chemical potential,
density, energy, entropy and pressure are computed within \abinit analogously to Eq.~\ref{eq:extfpmd_density}, using Fermi-Dirac integral
formulations~\cite{Blanchet2022}. The only input parameters required to control these
contributions are the number of KS orbitals \abivar{nband} and the flag
\abivar{useextfpmd} 1. The usage of this model is thus not more complex
than conventional KS-DFT since \abivar{nband} still works as a convergence
parameter, with its convergence being accelerated by extended contributions.

This implementation has been successfully tested via the
production of equation of state tables of multiple materials, including
hydrogen, boron, aluminum, and
iron~\cite{Blanchet2022, Blanchet2022b, Blanchet2022c, Blanchet2025}, as well
as for the computation of the ionization of carbon, and transport properties of gold and CH
mixtures~\cite{Clerouin2022, Stanek2024, Blanchet2024}, where the results were
compared with alternative methods when possible.
The implementation was tested in wide ranges of density and
from ambient temperature to ideal plasma temperatures ($T \approx 10^9\,$K).

\subsubsection{Thermal exchange-correlation functionals}
Recently, parallel progress has been made with the introduction of new
thermal exchange-correlation (xc) functionals that explicitly depend on the electronic
temperature, to capture complex effects in the warm
dense matter regime~\cite{Karasiev2014, 
Karasiev2016, Groth2017, Karasiev2018, Karasiev2022}.
\abinit\ was originally designed for
``$0\,$K xc functionals''.
In thermal
xc functionals, the xc entropy functional is included to yield the Helmholtz free xc energy 
\begin{equation}
  F_\mathrm{xc}[n] = E_\mathrm{xc}[n] - TS_\mathrm{xc}[n].
\end{equation}

To obtain the correct total internal energy and total entropy, both needed
for accurate equation of state data, $S_\mathrm{xc}$ must be accessible when
summing the energies after the free energy minimization. This
ensures that $S_\mathrm{xc}$ is properly taken into account, from the xc base
routine to the energy evaluation after the SCF cycle. This implementation is
compatible with both norm-conserving pseudopotentials and PAW decomposition.

The thermal xc functionals ``corr-KSDT''~\cite{Karasiev2014}
(based on LDA, \abivar{ixc} 51) and ``KDT16''~\cite{Karasiev2018} (based on GGA \abivar{ixc} 60)
were natively implemented alongside the previously implemented ``IIT'' thermal xc functional~\cite{Ichimaru1987} (based on LDA,
\abivar{ixc} 50). These functionals
are not yet compatible with spin-polarized calculations.

\abinit\ now also shares the electronic temperature with \textsc{LibXC} library~\cite{Lehtola2018}, enabling access to a wider range of thermal functionals.
For the moment \textsc{LibXC} provides the exchange-correlation free energy
of thermal functionals, it does not yet return the entropy contribution.
The \abinit\ interface is ready for future versions of \textsc{LibXC} which will return $S_\mathrm{xc}$.

\subsection{Use of meta-GGA exchange-correlation functionals}
\label{subsect:GS_metagga}

The term "meta-GGA" is used for the class of exchange-correlation functionals which include the electron kinetic energy density of the system $\tau({\bf{r}})$ \cite{Tao2003} in addition to its electron number density $n({\bf{r}})$ and possibly its gradient and Laplacian.
Here, $n({\bf{r}})$ and $\tau({\bf{r}})$ are determined self-consistently by summing over all occupied states, using a similar notation to that in Eq. (\ref{eq:extfpmd_density}) at $T=0K$.
\begin{equation}{\label{eq:nandtau}}
n({\bf{r}})\equiv\sum_{n \kk}^{(occ)} w_{n\kk} |\psi_{n,\kk}({\bf{r}})|^2\;\;\;\tau({\bf{r}})\equiv\frac{1}{2}\sum_{n \kk}^{(occ)} w_{n\kk} |\nabla\psi_{n,\kk}({\bf{r}})|^2   
\end{equation}
Here $w_{n\kk}$ denotes the state occupancy times the Brillouin zone weighting factor. The general form of the meta-GGA exchange-correlation energy $E_{xc}$ is given in terms of the spacial integral of the kernel function $f_{xc}$
\begin{equation}{\label{eq:mggaxcdef}}
E_{xc}\equiv\int d^3r f_{xc}(n({\bf{r}}),|\nabla n({\bf{r}})|,\Delta n({\bf{r}}),\tau({\bf{r}})).
\end{equation}
Note that these equations represent a spin unpolarized case; \abinit can also similarly treat spin polarized systems within meta-GGA.  

Although earlier versions of \abinit had some meta-GGA capabilities, the new implementation extends to the Projector Augmented Wave (PAW) methodology of Bl\"{o}chl \cite{Bloechl1994}, notably with functionals from the SCAN family (SCAN, 
rSCAN, r$^2$SCAN)~\cite{Furness2020,Sun2016}. 
 Within the PAW framework, all varieties of meta-GGA functionals are available, including those employing the Laplacian of the density $(\Delta n({\bf{r}}))$ and those based on kinetic energy density $(\tau({\bf{r}}))$. 
 For the latter, inspiration was drawn from the work of Sun \textit{et al.}~\cite{Sun2011}. Functionals referred to as \textit{potential-only}, such as the Becke-Johnson functional~\cite{Becke2006}, are also accessible. 
 When the exchange-correlation functional includes the kinetic energy density $(\tau({\bf{r}}))$, the Kohn-Sham Hamiltonian equations
are approximated as "generalized" Kohn-Sham equations \cite{Yang2016} in that the effective exchange-correlation
"potential" includes gradient operators of the form:
\begin{multline}
  \label{eq_mgga_potential}
  v^{\rm meta-GGA}_{xc}(\mathbf{r}) =
    \frac{\partial \left(f_{xc} \right)}{\partial n}
  - \nabla \cdot \left(\frac{\partial f_{xc}}{\partial |\nabla n|}\frac{\nabla n}{|\nabla n|} \right)\\
  + \Delta \left(\frac{\partial \left(f_{xc} \right)}{\partial \Delta n} \right)
  - \frac{1}{2} \nabla \cdot \left( v_{\tau}(\mathbf{r}) \nabla \right),
\end{multline}
where
\begin{equation}{\label{eq:vtaudef}}
v_{\tau}(\mathbf{r})= \frac{\partial f_{xc}}{\partial \tau}.
\end{equation}

The PAW formalism \cite{Bloechl1994} evaluates the effective Hamiltonian of the system, including the exchange-correlation contribution given in Eq. (\ref{eq_mgga_potential}) with two contributions.  One contribution is from valence electrons, evaluated within the plane-wave basis with a pseudized version of the effective Hamiltonian. The other contribution is from a sum of one-center integrals within non-overlapping "augmentation" spheres evaluated using two types of atomic basis functions -- one representing the all electron system and the other, consistently representing the pseudo system. These one-center integrals are evaluated efficiently using angular and radial grids.  The details of the PAW implementation in \abinit are described by Torrent and collaborators \cite{Torrent2008}. In order to adapt \abinit for accurate treatment of  meta-GGA, the code  modifications  were straight forward, mainly concerning the form of the last term of Eq. (\ref{eq_mgga_potential}).    In addition to evaluating the Hamiltonian, the meta-GGA treatment also affects the expressions for the forces and stresses.   These expressions extend the earlier work of Dalcorso and Resta \cite{Dalcorso1994}  as described by Sun and collaborators \cite{Sun2011} and further work by Charraud in his Ph. D. thesis \cite{Charraud2021}.

In the \abinit code, access to the meta-GGA exchange-correlation functionals is obtained from the \textsc{LibXC} library \cite{Marques2012,Lehtola2018}, using \abinit's \abivar{ixc} keyword which is related to id labels in the \textsc{LibXC} repository. The \abivar{ixc} variable takes negative values for 
\textsc{LibXC} funcionals based on the numbering scheme of the library.   For
instance, to conduct calculations using SCAN, set \abivar{ixc}$=-263267$; 
for the r$^2$SCAN functional, set \abivar{ixc}$=-497498$. Experience shows that, as for self-consistent atomic calculations \cite{Holzwarth2022},  the meta-GGA functional forms sometimes have numerical sensitivities so that adjustment to the self-consistency and/or optimization algorithms may be needed.

Atomic datasets for PAW calculations with meta-GGA functionals are available with an updated version of \textsc{atompaw} \cite{Holzwarth2022}. It is also possible to use \abinit to perform meta-GGA calculations using non meta-GGA atomic datasets generated by the \textsc{atompaw} code: the output atomic datasets in xml format always include all-electron and pseudo core kinetic energy densities on a radial grid. These atomic datasets can be used in \abinit to approximate meta-GGA calculations by adjusting the \abivar{ixc} value. Preliminary experience shows that results can be quite similar to those obtained using consistent meta-GGA datasets. Further investigation on this topic is ongoing.   

In addition to implementation of meta-GGA with the  PAW formalism,  progress has also been made with meta-GGA implementations within traditional pseudopotential methodologies. 
The pseudopotential generation software ONCVPSP~\cite{Hamann2013} now provides a development version METAPSP\cite{metapsp} with norm-conserving meta-GGA potentials compatible with \abinit. The UPF2 file format is used and has been augmented with fields for the isolated atom (\texttt{PP\_TAUATOM}) and model-core (\texttt{PP\_TAUMOD}) kinetic energy densities, in coordination with the developers of the QuantumEspresso/UPF toolkit. The former field will be used in the future to initialize the SCF cycle and improve convergence, while the latter is used to complete the expressions of total energy derivatives. Both are read by \abinit, which then follows the same routines as for the PAW implementation above to perform ``full'' meta-GGA calculations with core and valence electrons on the same xc footing.

\subsection{Orbital magnetism and magnetic shielding}
\label{subsect:GS_NMR}

Calculation of the orbital magnetism has been implemented in \abinit\ 
in two ways: a complete treatment particularly suited for the 
magnetism in a diamagnetic insulator induced by a nuclear magnetic
dipole moment, and a simpler treatment within atom-centered spheres,
well-suited for antiferromagnetic materials.

The complete treatment is outlined in Ref.~\cite{Zwanziger2023}.  The implementation is accomplished 
by evaluating the first order expansion of the energy  with respect to a 
magnetic field,  which follows because the magnetization component $M_i$ is given by 
$M_i=-(1/\Omega)(\partial E/\partial B_i)$, where 
$\Omega$ is the cell 
volume. The expansion involves a number of technicalities arising from
the gauge variance of the vector potential as well as the periodic 
boundary conditions; these were resolved using magnetic translation 
invariance~\cite{Essin2010a} as detailed in 
Refs.~\citenum{Gonze2011a,Zwanziger2023}.
The resulting expression for the first order perturbed energy ($E^{(1)}$) in the full PAW context involves 
only the occupied ground-state wavefunctions and their derivatives with 
respect to $\kk$-points, projected on the unoccupied subspace.  The 
response $E^{(1)}$ is then assembled from these components, as detailed 
in~\cite{Zwanziger2023}. Calculation of the response is triggered 
with the variable \abivar{orbmag}.

The necessary derivative wavefunctions may be obtained in two ways,
either from a DFPT calculation, using \abivar{rfddk}, or by finite
differences between neighboring $\kk$ points, using \abivar{berryopt} set to -2. 
Typically the DFPT approach converges much faster with $\kk$-mesh
density. Both methods are compatible with a variety of exchange and 
correlation functionals, including LDA, GGA, and metaGGA using kinetic 
energy density, as well as with electronic spin degrees of freedom.

The \abivar{orbmag} calculation is designed to work together with a nuclear 
magnetic dipole moment $\mathbf{m}$, input to \abinit with \abivar{nucdipmom}, to produce the magnetic shielding $\sigma$ as measured in NMR 
experiments. The shielding  is a joint response to magnetic field and 
nuclear dipole, such that
\begin{equation}
  \sigma_{ij}=-\frac{\partial^2 E}{\partial m_i\partial B_j} = \Omega\frac{\partial M_j}{\partial m_i}.
\end{equation}
This approach to the shielding in terms of the magnetization $M_j$ is 
termed the converse method~\cite{Thonhauser2009a,Ceresoli2010b}. 
In insulators, the magnetization in 
the absence of a dipole is always zero, so the shielding can be read off 
directly from the magnetization in the presence of a small nuclear 
magnetic dipole. Our approach and numerous examples are detailed in 
Ref.~\citenum{Zwanziger2023}, and a tutorial has been added to the
\abinit web site.

\begin{figure}
\includegraphics[width=82mm]{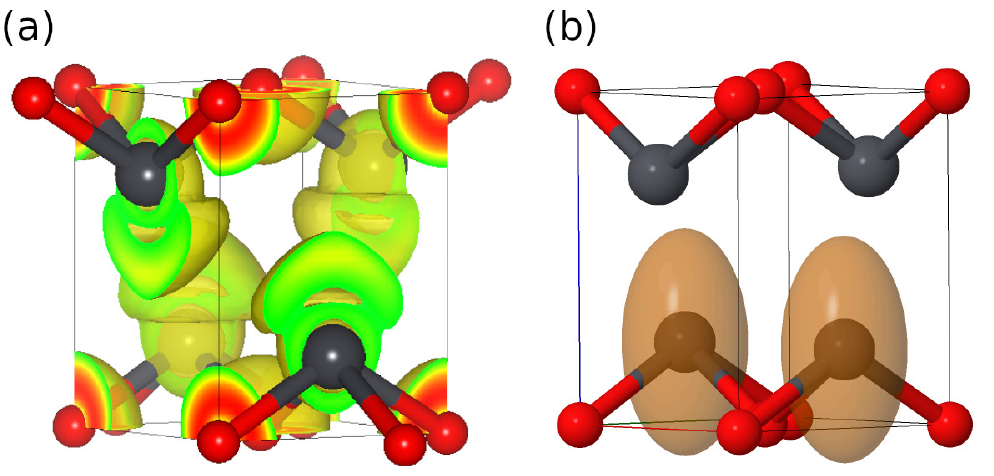}
\caption{\label{litharge}(a) Electron localization function for PbO in the litharge structure, showing the lone pairs on lead (grey spheres). (b) In the same structure, the magnetic shielding tensor orientations represented as ellipsoids, with principal diameters proportional to the principal shielding values, computed using ZORA corrections and the
r$^2$SCAN functional. The lone pairs clearly play the major role in the magnetic shielding.}
\end{figure}

Furthermore, we have implemented the zeroth order regular approximation 
(ZORA) terms for the nuclear dipole electron spin interaction through the 
\abivar{pawspnorb} key word~\cite{Zwanziger2025a}; this addition gives 
excellent shielding values in the presence of spin-orbit coupling. Fig.~\ref{litharge} shows a result for the litharge polymorph of lead oxide, a system with a pronounced lone pair on lead and significant relativistic effects.

The complementary atom-centered approach has been implemented by integrating the orbital magnetization inside the PAW spheres only (\abivar{prt\_lorbmag} input flag).
This approach is useful for antiferromagnetic cases where the total orbital magnetization is zero but not the local one on each atom (see, {\em e.g.}, Ref.~\cite{braun2024}). 
Care should be taken for systems where the interstitial contribution between spheres is  important~\cite{ceresoli2010a}.


%% file: 3_DFPT_EPC.tex
\section{Responses to perturbations and related properties}
\label{sect:DFPT}


The linear and non-linear responses of materials to different applied static or dynamic fields  (electric field, magnetic field), as well as to strain, play an important role in many practical applications. Such properties can be obtained, {\em e.g.}, from density-functional perturbation theory (DFPT)~\cite{Baroni1987, Gonze1989, Gonze1995, Gonze1997, Gonze1997a, Baroni2001, Gonze2024}, that also yields vibrational and related properties of materials, like electron-phonon coupling.
New physical properties based on such responses and new perturbations have been introduced in \abinit in the last five years.

\subsection{Spatial dispersion}
\label{subsect:DFPT_spatial}

Spatial dispersion refers to the dependence of a given linear-response property on the momentum at which it is probed.
Equivalently, it can be understood as the response of the crystal to a real-space gradient of the applied field.
Notable manifestations in condensed matter are natural optical activity
(NOA), describing the ability of certain crystals of rotating the plane of
polarization of transmitted light, and flexoelectricity, describing the macroscopic polarization response to a strain gradient.
Apart from the outstanding practical relevance (for example, NOA is important for industry as a characterization tool for chiral molecules), this class of properties has a strong fundamental interest because of its intriguing connections to quantum geometry~\cite{PhysRevB.82.245118,10.21468/SciPostPhys.14.5.118}.
In the electrical case, there are obvious formal links to orbital magnetism, too; and indeed, many crystal properties that emerge under an applied magnetic field ({\em e.g.}, Lorentz forces, rotational $g$-factors, {\em etc.}) can be captured via the same methodologies as conventional spatial dispersion effects.

\abinit has been at the forefront of these developments in the past ten years or so, and continues to lead in this field. Our pioneering implementation~\cite{Royo2019} of dynamical quadrupoles and clamped-ion flexoelectric coefficients (described in our earlier report) has now been complemented by the incorporation of lattice-mediated flexoelectric coefficients~\cite{Royo2022},
geometric magnetization (relevant for Lorentz forces and rotational 
$g$-factors)~\cite{Zabalo2022},  and NOA~\cite{Zabalo2023}.
The latter two properties required the implementation of a new perturbation, that is, the first-order wavefunction response ---specifically its orbital part--- to a homogeneous magnetic field~\cite{Zabalo2023} which
can be useful in other contexts as well.

Of special note, all spatial dispersion properties can now be calculated either with LDA or GGA: this feature was introduced in Ref.~\citenum{Springolo2021}, but never properly documented.
These calculations currently require the use of norm-conserving pseudopotentials. Support for pseudopotentials with exchange-correlation nonlinear core corrections has recently been implemented, enabling the calculation of the dynamical quadrupoles and the NOA tensor. However, flexoelectric coefficients can still only be computed using norm-conserving pseudopotentials without such core corrections.
Several capabilities of the long-wave module have been applied to both three-dimensional and two-dimensional crystals, {\em e.g.} in the
cases of dynamical quadrupoles~\cite{Royo2021} 
and flexoelectricity~\cite{PhysRevLett.131.236203,Springolo2021}.
Note that the specifics of 2D materials ({\em e.g.} the electrical and mechanical boundary conditions) can have a
rather strong impact on the calculated coefficients, in some cases even affecting their very definition;
we refer to the original publications~\cite{Royo2021, Springolo2021, PhysRevLett.131.236203} for further details.


\subsection{Phonon angular momentum}
\label{subsect:DFPT_phonon_angmom}

Phonons can possess nonzero angular momentum in systems lacking spatial inversion symmetry~\cite{Coh2023}. The Anaddb utility of \abinit is now able to compute phonon angular momentum. The angular momentum for one mode $\nu$ with wave-vector $\qq$ is given by:
\begin{equation}
    l_{\qq\nu\alpha} = -i\hbar \sum_{\kappa} \varepsilon_{\alpha\beta\gamma} \varepsilon_{\qq\nu\kappa\beta}^* \varepsilon_{\qq\nu\kappa\gamma},
\end{equation}
where $\varepsilon_{\alpha\beta\gamma}$ is the Levi-Civita symbol, and $\varepsilon_{\qq\nu\kappa\alpha}$ is the displacement of atom $\kappa$ in direction $\alpha$ for eigenmode $\qq \nu$~\cite{Zhang2014}. The phonon angular momentum is automatically computed if a Fourier interpolation and band structure are requested. $l_{\qq\nu\alpha}$ is expressed in units of $\hbar$ and in Cartesian coordinates $\alpha$, for each phonon mode $\nu$ at each $\qq$-point. It is written to output in both text and NetCDF formats (in the \abioutfile{PHANGMOM} and \abioutfile{PHBST.nc} files respectively). The \abioutfile{PHBST.nc} file can be used to plot the angular momentum with the \abipy API.

\subsection{Electron-phonon interaction for transport properties}
\label{subsect:DFPT_transport}
Electron scattering by phonons is the main intrinsic mechanism limiting transport in solids. The standard framework for first principles transport is the Boltzmann transport equation (BTE).
In the hypothesis of weak external fields, the out-of-equilibrium statistical distribution function $f$ can be expressed as a small correction to the equilibrium Fermi-Dirac occupation function according to:
\begin{equation}\label{eq:fermi}
f_\nk(\mu, T) = f^0(\ee_\nk, \mu, T) + \delta f_\nk(\mu, T)
\end{equation}
where the unknown $\delta f_\nk$ term is linear in the external fields and $f^0(\ee_\nk, \mu, T)$ is the local equilibrium (Fermi-Dirac) distribution. The linearized version of the BTE is expressed as:
\begin{multline}
\label{eq:linearizedBTE}
\PDER{f^0}{\ee_{\nk}} \vv_\nk \cdot \biggl[ 
-e \biggl(\EE + \dfrac{1}{e} \PDER{\mu}{\rr} \biggr) 
- \dfrac{(\ee_\nk - \mu)}{T} \PDER{T}{\rr} 
\biggr] \\
+ \vv_\nk \cdot \PDER{\delta f_\nk}{\rr}
= \mcL_{n\kk}[f^0, \delta f]
\end{multline}
%
where 
the right hand side is the scattering integral, a functional of $f$ that represents the net number of particles entering/leaving the infinitesimal phase space region around $(\rr, \kk)$ due to scattering processes. This equation forms the basis for both the relaxation-time approximation (RTA) and the iterative Boltzmann transport equation (IBTE) formalisms. In both approaches, the resulting expression for $\delta f_\nk$ is inserted into the current-density formulas to obtain the full set of transport tensors, including electrical conductivity, resistivity, the Seebeck and Peltier coefficients, and the electronic thermal conductivity.\\

The electron-phonon (EPH) module of \abinit provides a complete framework for computing phonon-limited transport properties. The electron-phonon self-energy (Fan–Migdal + Debye–Waller) \cite{Giustino2007} is evaluated using first-principles Kohn–Sham (KS) wavefunctions and a Fourier interpolation of the DFPT scattering potentials. This interpolation yields a much denser $\qq$-point sampling than that employed in the DFPT calculation \cite{Brunin2020, Brunin2020b}. 
For transport applications, a preliminary run with \abivar{eph\_task -4} computes the imaginary part of the Fan–Migdal self-energy at the Kohn–Sham energy. The calculation employs the tetrahedron method and KS wavefunctions restricted to an energy window around the band edge. The resulting mode-resolved linewidths, carrier lifetimes, and interpolated e–ph matrix elements are written to the file \abioutfile{SIGEPH.nc}, which is subsequently read by all transport solvers. When strong electron-phonon coupling invalidates the quasi-particle picture, a full frequency-dependent self-energy—including both real and imaginary parts—must be evaluated with \abivar{eph\_task 4}. \textsc{Abinit} can then post-process these data to compute transport properties at the various levels of theory outlined in the following sections.

\subsubsection{RTA under thermal and electrical gradients}
Within the RTA, the scattering integral $\mcL_{n\kk}$ in Eq.~\ref{eq:linearizedBTE} is replaced by: $\mcL_{n\kk}[f^0, \delta f] =  -\dfrac{\delta f_\nk}{\tau_\nk}$.
%
%
The RTA simplifies the original problem by making scattering depend solely on the relaxation time $\tau_\nk$. The evaluation of $\tau_\nk$ distinguishes the momentum-relaxation-time approximation (MRTA) from the self-energy relaxation-time approximation (SERTA): SERTA neglects most back-scattering processes, whereas MRTA includes them.  In the new version of \abinit, the RTA transport tensors are calculated considering a chemical potential that varies with temperature. In previous versions, the chemical potential was considered constant and equal to the ground state Fermi level. Transport coefficients within the RTA~\cite{Giustino2007} are computed using \abivar{eph\_task 7}, which reads the carrier lifetimes from the file \abioutfile{SIGEPH.nc}.

\subsubsection{IBTE under thermal and electrical gradients}
The iterative BTE solution~\cite{Claes2022, Willis2023} goes beyond the RTA by explicitly including in and out electron-phonon scattering. 
The right-hand side of Eq.~\ref{eq:linearizedBTE} becomes $-\dfrac{\delta f_\nk}{\tau^0_\nk} + \mathcal{L}^{\text{e-ph}}_{n\kk}[f^0, \delta f]$, with the first term equivalent to the RTA formalism and the second term explicitly taking into account the linearized electron-phonon scattering integral. 
We also express the correction to the distribution function (Eq.~\ref{eq:fermi}) as a first contribution linear in $\EE$ and a second contribution linear in $\TT$, 
such that: $\delta f_\nk = \FF^{\mathbf{\mathcal{E}}}_{\nk} \cdot \EE + \FF^T_\nk \cdot \boldsymbol{\nabla}T / T $, where $\FF^T_\nk$ and $\FF^{\mathbf{\mathcal{E}}}_\nk$ are tensors giving the first derivatives of the distribution function $f_\nk$ with respect to $\boldsymbol{\nabla}T$ and $\EE$, respectively. 
These tensors are calculated using an iterative scheme which yields a solution for $\delta f_\nk$. 
Then, the transport tensors can be calculated within IBTE. 
In the previous versions of \abinit, only calculations of conductivity and resistivity were available. 
The Seebeck coefficient, the electronic thermal conductivity, and the Peltier coefficient are now implemented as well. 
A more detailed explanation of the formalisms and explicit mathematical derivations can be found in Ref.~\cite{Allemand2025}.
This calculation is invoked with \abivar{eph\_task 8} and uses collision terms stored in \abioutfile{SIGEPH.nc}.
\subsubsection{Kubo-Greenwood formalism}
It is important to recognize that both the RTA and IBTE frameworks rely on well-defined quasi-particles with spectral functions $A_\nk(\omega)$ sharply peaked around the QP energy. However, in systems where strong electron–phonon coupling redistributes spectral weight and generates polaronic sidebands, this assumption fails, and a rigorous description must therefore go beyond the BTE by using the Kubo–Greenwood formalism, in which the static conductivity is given by:
%
\begin{eqnarray}
\sigma  &=& \dfrac{\pi e^2}{\Omega N_\kk} \sum_{\kk n}
\vv_{n\kk} \otimes \vv_{m\kk} 
\nonumber
\\
&&\int A_\nk(\ww')A_\nk(\ww'+\ww)
\dfrac{f(\ww') - f(\ww'+\ww)}{\ww}
\dd\ww',
\nonumber
\\
\end{eqnarray}
where $\vv_{n\kk}$ are the matrix elements of the velocity operator, and $A_\nk(\ww')$ the spectral function computed from the imaginary part of the retarded Green's function.
The spectral function can be obtained either by solving the Dyson equation for $G$ with the standard e–ph self-energy in the Migdal approximation (yielding the \emph{Dyson–Migdal} spectral function), or by employing more advanced approaches, such as the cumulant-expansion method, which captures satellite features more accurately \cite{Nery2018, Abreu2022, Zhou2019}. Once $\sigma$ is known, auxiliary quantities such as carrier mobility can also be evaluated. To enable direct comparison and a clearer understanding of spectral-function effects, the current \textsc{Abinit} implementation provides Kubo–Greenwood transport coefficients computed with both Dyson–Migdal and cumulant spectral functions, accessible via \abivar{eph\_task 9}. However, these calculations are considerably more demanding because they must resolve the full frequency dependence of the e–ph self-energy, as mentioned earlier, whose convergence requires extensive summations over empty electronic states. \abinit can mitigate this cost by approximating high-energy contributions through a non-self-consistent solution of the Sternheimer equation, activated with \abivar{eph\_stern}~\cite{Abreu2022}. In the Sternheimer approach, the e–ph self-energy is approximated by an adiabatic expression in which the phonon frequencies are neglected; this approximation remains valid provided that a sufficiently large number of bands above the states of interest are explicitly included.


\subsection{Electron-phonon interaction for polarons and for band-gap renormalization}
\label{subsect:DFPT_polarons}

Electron-phonon interaction in materials may lead to polaron formation:
a particle couples to the lattice vibrations, becoming associated with the phonon cloud.
Two limiting regimes of polaronic effects are typically distinguished: weak and strong coupling.
In the weak-coupling regime, the particle coherently couples to phonons; in the strong-coupling regime, it becomes self-trapped in the induced deformation field.
In \abinit~10, the EPH code can treat both regimes for electrons and holes.

In the weak-coupling limit, zero-point renormalization of the gap (ZPR) can be computed using the generalized Fr\"ohlich (gFr) model~\cite{Miglio2020} (\abivar{eph\_task} 6):
\begin{multline}
    \mathrm{ZPR}^\mathrm{gFr}
    =
    \sigma
    \sum_{\nu n}
    \frac{1}{\sqrt{2}\Omega_0 n_\mathrm{deg}}
    \int_{4\pi} d\hat{\mathbf{q}}
    \left( m^*_n({ \hat{\mathbf{q}} })\right)^{1/2} \\
    \times
    \left( \omega_{\nu,\mathrm{LO}}({ \hat{\mathbf{q}} })\right)^{-3/2}
    \left(
    \frac{\hat{\mathbf{q}} \cdot \mathbf{p}_{\nu}({ \hat{\mathbf{q}} })}{\epsilon^\infty ({ \hat{\mathbf{q}} }) }
    \right)^2,
\end{multline}
where $\sigma = \pm 1$ for VBM/CBM, $\Omega_0$ is the primitive cell volume, and effective masses $m^*_n$, phonons $\omega_{\nu,\mathrm{LO}}$, mode polarities $\mathbf{p}_{\nu}$ and dielectric tensor $\epsilon^\infty$ are direction dependent at $\hat{\mathbf{q}} \to \Gamma$.
Summation runs over multiple LO phonon modes and $n_\mathrm{deg}$ degenerate bands.

For cubic systems, polaron effective mass enhancement is accessible at the same level of perturbation theory~\cite{Guster2021} (\abivar{eph\_task} 10).
Both calculations require a few macroscopic input parameters, obtained from the ground-state and DFPT modules.

The gFr model is a cheap alternative to the full first-principles computation of the ZPR, as previously implemented in ABINIT~\cite{Gonze2016} and recently applied to the spin-orbit case~\cite{Brousseau-Couture2023} and for high-throughput studies~\cite{deMelo2023}, the latter including comparison with the gFr model.


In the strong-coupling limit, self-trapped polarons can be computed using the variational polaron equations (VarPEq) framework~\cite{Vasilchenko2022} (\abivar{eph\_task}~13), based on the reciprocal-space formalism for polarons~\cite{sio_ab_2019, sio_polarons_2019}.
Here, a self-trapped polaron is characterized by its binding energy, defined as the total energy difference between polaronic and pristine geometries of a system with $N \pm 1$ electrons:
\begin{equation}
    E_\mathrm{pol} =
    E \left( N \pm 1, \Delta \boldsymbol{\tau} \right)
    -
    E \left( N \pm 1, \Delta \boldsymbol{\tau} \equiv 0 \right),
\end{equation}
where $\Delta \boldsymbol{\tau}$ denotes the collective lattice distortion.
VarPEq formalism provides a variational expression for binding energy
\begin{multline}\label{eq:varpeq}
    E_\mathrm{pol}
    \left( \boldsymbol{A}, \boldsymbol{B} \right)
    = \frac{1}{N_p} \sum_{n\mathbf{k}}
    | A_{n\mathbf{k}} |^2
    \varepsilon_{n\mathbf{k}}
    +
    \frac{1}{N_p} \sum_{\mathbf{q}\nu}
    | B_{\mathbf{q}\nu} |^2 \omega_{\mathbf{q}\nu} \\
    -  \frac{1}{N_p^2} \sum_{mn\nu \mathbf{kq}}
     A^*_{m\mathbf{\mathbf{k+q}}} B^*_{\mathbf{q}\nu} g_{mn\nu}(\mathbf{k, q})
    A_{n\mathbf{\mathbf{k}}} + \mathrm{(c.c)},
\end{multline}
where electronic bands $\varepsilon_{n\mathbf{k}}$, phonons $\omega_{\mathbf{q}\nu}$ and electron-phonon coupling $g_{mn\nu}(\mathbf{k, q})$ act as input, and $\boldsymbol{A}$ and $\boldsymbol{B}$ are the variational coefficients.

\begin{figure}
    \centering
    \includegraphics[width=1.\linewidth]{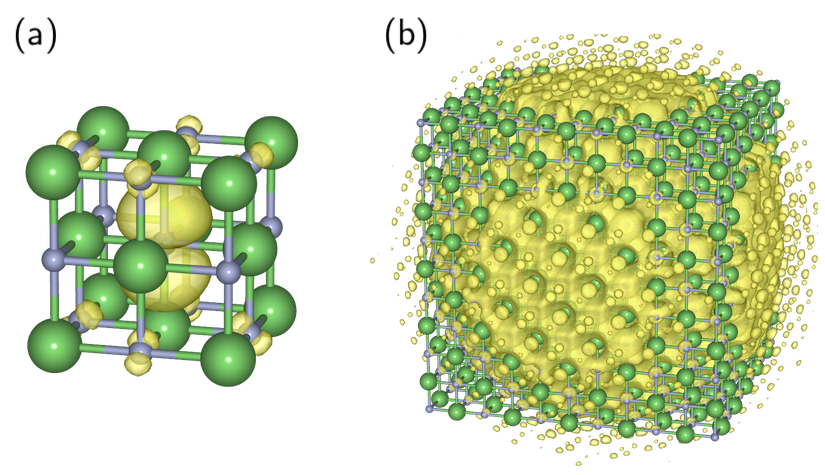}
    \caption{
    Polaron charge distributions in LiF, obtained with the variational polaron equations framework.
    The panels correspond to: (a) a small hole and (b) a large electron polaron in LiF.
    }
    \label{fig:polarons}
\end{figure}

The parametrization of Eq.~\ref{eq:varpeq} requires the computation of \abioutfile{WFK}, \abioutfile{DDB}, \abioutfile{DVDB}, and \abioutfile{GSTORE.nc} files.
Then, it can be optimized using a preconditioned conjugate gradient solver, which enables the search for multiple polaronic states within a single system, typical in materials with degenerate VBM/CBM~\cite{vasilchenko_polarons_cubic_2024}.
It has been successfully applied to compute degenerate polaronic states 
from first principles
~\cite{Vasilchenko2025arxiv}.
Fig.~\ref{fig:polarons} demonstrates polarons in LiF obtained via the VarPEq formalism.

For post-processing, VarPEq calculations produce \abioutfile{VPQ.nc} files, which can be used to compute the polaron charge density $\rho(\mathbf{r})$ and lattice distortion $\Delta \boldsymbol{\tau}$ in a hosting supercell (\abivar{eph\_task}~$-13$).
Moreover, \abipy provides a toolkit for the \abioutfile{VPQ.nc} data analysis.
Practical examples demonstrating the VarPEq workflow are available on the \abinit website~\cite{abinit_website}.
In particular, a tutorial presenting a detailed, step-by-step procedure for computing both small and large polarons, has been added to the official documentation.

\subsection{Improvements in the DDB - derivative database file}
\label{subsect:DFPT_ddb}

The derivative database file (DDB) produced by \abinit contains different types of energy derivatives with respect to perturbations. The perturbations include atomic displacements, strain, electric field, and magnetic field, while the derivatives may be of 1st order (forces, stress), of 2nd order (dynamical matrix, Born effective charge tensor, etc.), or 3rd order (quadrupole tensor, etc.).

In \abinit 10, a new NetCDF format has been designed for the derivative database (\abioutfile{DDB.nc}), which is used when the user requests the NetCDF output format through the input variable \abivar{iomode}, or when it is set as the default output format at compilation time. While the plain text format is still available for the DDB, the NetCDF format includes additional meta-information, such as the input file that produced it, more information on the pseudopotentials, and a literal description of the perturbations. 

The \abiexec{mrgddb} utility has also been refactored. Its main purpose is to merge several DDB files into a single DDB, so that the computation of a complete derivative tensor may be split into distinct calculations for each row. In \abinit 10, the \abiexec{mrgddb} utility can also convert a single file between plain text and NetCDF format.

Finally, other types of derivatives have been implemented or unified in the DDB format, such as second derivatives of the electronic eigenvalues with respect to atomic displacements (\abioutfile{EIGR2D.nc}) and the Berry curvature (\abioutfile{BERRY\_DDB.nc}).

\subsection{Optical conductivity with spin-orbit coupling}
\label{subsect:DFPT_optics_SOC}

Transport properties, such as electrical conductivity, dielectric functions, reflectivity, refractive index, and absorption coefficient, among others, are of significant importance among the experimental observables related to electronic properties.

In the context of DFT, computations are executed within a \textit{single-particle} framework: an electron transitions from a core or valence orbital state to an excited state, and the transition energy corresponds to the energy difference between the two states, neglecting many-body effects. Within this framework, the real part of the electrical conductivity $\sigma_1$ is obtained through the Chester-Thellung formulation~\cite{Chester1959}, revised by Kubo-Greenwood~\cite{Kubo1957,Greenwood1958}, which is expressed in terms of the velocity operator $\mathbf{v}$:
\begin{multline}
 \label{eq_localized_transport_kubogreenwood_2}
 \sigma_1(\mathbf{k},\omega)=\frac{2\pi}{3\omega\Omega}\sum_{mm'} \left(f_{m\mathbf{k}}-f_{m'\mathbf{k}}\right)
 \left\vert
 \langle \Psi_{m\mathbf{k}} | \mathbf{v} | \Psi_{m'\mathbf{k}} \rangle
 \right\vert^2 \\
  \delta(\epsilon_{m\mathbf{k}}-\epsilon_{m'\mathbf{k}}-\omega),
\end{multline}
where $\mathbf{k}$ is the wave vector, $\omega$ the energy of the photon and $\Omega$ the volume of the unit cell. $f_{m\mathbf{k}}$ is the Fermi-Dirac occupation factor of the band $m$, corresponding to the eigenenergy $\epsilon_{m\mathbf{k}}$.
\noindent The total electrical conductivity $\sigma_1(\omega)$ is obtained through a direct summation over all the wave vectors $\mathbf{k}$.

In a non-relativistic approach, the velocity operator simply reduces to $\mathbf{v}=-i\boldsymbol{\nabla}$. In the relativistic case, it includes a contribution from spin-orbit coupling (incorporating the fine structure constant $\alpha$, Pauli matrices $\boldsymbol{\sigma}$, and the electronic potential $V({\mathbf{r}}$):
\begin{equation}
 \label{eq_localized_transport_velocity_soc}
\mathbf{v}=-i\boldsymbol{\nabla}+\frac{1}{4 m^2_e c^2} \boldsymbol{\sigma}\times\nabla V({\rr}).
\end{equation}
Regarding core spectroscopy (XANES), one of the two states $\Psi_{m\mathbf{k}}$ involved in the electronic transition is a core state $\Phi_c$ with energy $\epsilon_c$, whose extent is limited to the localized region near an atom.

In Ref.~\citenum{Mazevet2010}, the expression for electrical conductivity $\sigma_1$ has been formulated within the projector augmented-wave (PAW) approach. This reformulation thus modifies the final expression for the absorption cross-section for the core state $\Phi_c$ with energy $\epsilon_c$ as~\cite{Brouwer2021}:
\begin{multline}
 \label{eq_localized_transport_xanes_1}
 \sigma_1(\mathbf{k},\omega)=\frac{2\pi}{3\omega\Omega}\sum_{m}\left(1-f_{m\mathbf{k}}\right) \delta(\epsilon_{c}-\epsilon_{m\mathbf{k}}-\omega) \\
 \left\vert
 \sum_{R,i}
 \langle \Psi_{m\mathbf{k}} | \tilde{p}^{R}_{i} \rangle
 \langle \phi^{R}_{i} | \mathbf{v} | \Phi_c \rangle
 \right\vert^2,
\end{multline}
where $\phi^{R}_{i}$ is an element of the local PAW basis and $\tilde{p}^{R}_{i}$ is the associated projector.

The primary advantage of this approach is that it does not require the use of specific atomic data ({\em i.e.}, pseudopotentials), including relativistic basis functions. However, it is essential to have access to the core orbitals $\Phi_c$, which, in a relativistic approach, are two-component spinor functions, solutions to the Dirac equation. We have adapted the \textsc{atompaw} software~\cite{atompaw_website} to produce PAW atomic datasets that include relativistic core orbitals. These are expressed using relativistic spherical harmonics~\cite{Martin2004}.
All the electronic transport properties are accessible using the executable \abiexec{conducti}, available in the \abinit package. \abiexec{conducti} uses $\langle \Psi_{m\mathbf{k}} | \mathbf{v} | \Psi_{m'\mathbf{k}} \rangle$ transition elements, which are produced by \abinit using the \abivar{prtnabla} keyword. Spin-orbit coupling is activated, as usual, with the \abivar{pawspinorb} keyword.

%% file: 4_excitationsandcorrelations.tex
\section{Excited Electronic States and Correlations}
\label{sect:excitations}


This section gathers numerous advances beyond
ground-state DFT that have been implemented in \abinit{} in the last four years.
These extensions to ground-state DFT not only 
allow us to calculate the observables of spectroscopic experiments
but incidentally have consequences on ground-state quantities.

Real-time TDDFT, dynamical mean-field theory (DMFT), and the $GW$ approximation in the Green's function theory typically address optical and charged excitations. However, DMFT and $GW$ give access to ground-state quantities, such as the correlation energy or the one-body reduced density-matrix.

Since at the very heart of our coding we constantly employ the Coulomb integrals, it is a direct consequence to export those integrals so that they could be used by external codes. This capability is demonstrated with the linkage to the coupled-cluster code CC4S~\cite{cc4s_website}.

\subsection{Real-time time-dependent DFT}
\label{subsect:excitations_realtime}

Time-dependent density functional theory (TDDFT) is a well-known method to investigate 
the electronic response of materials to \emph{time-dependent} perturbations~\cite{Runge1984,Ullrich2012}.
This method is usually formulated in frequency space within linear response theory 
leading to the so-called linear response TDDFT method which is extensively 
used to study excited states properties of molecular systems and materials.
Alternatively, the time-dependent Kohn-Sham equations can be numerically integrated 
in \emph{real-time} leading to the so-called real-time TDDFT (RT-TDDFT) method, 
which gives direct access to the electron dynamics.
Unlike its linear response counterpart, RT-TDDFT is not limited to the linear response 
regime and can thus be used to evaluate the response to strong perturbations, 
such as intense laser pulses~\cite{Yabana2012}. 
Furthermore, the electron dynamics can be coupled with molecular dynamics on 
the ions, opening the door to non-adiabatic dynamics.
 
Real-time TDDFT is now being implemented in \abinit and can be used by setting
the variable \abivar{optdriver} to 9. Additional parameters, such as the electronic time
step \abivar{dtele} and the number of propagation steps \abivar{ntime},
need to be specified. A complete list of variables is available in the documentation~\cite{abinit_website}.
The implementation in \abinit can be used in particular to study the 
response to an external impulse electric field.
This is done in the long wavelength approximation for which the electric field is
considered constant in space and using the \emph{velocity gauge} so that 
the electric field is only described via the vector potential $\mathbf{A}(t)$.
Within this framework, only the kinetic part of the Hamiltonian is modified via 
the momentum operator that is now given by $-i\boldsymbol{\nabla} + \mathbf{A}(t)$.
This is activated by setting the variable \abivar{td\_ef\_type} to 1.
Note that the application of an external electric field is only possible within the 
projector augmented wave (PAW) approach~\cite{Torrent2008} for now.
The response to the electric field can be evaluated by computing the induced 
macroscopic current density
 \begin{equation}
    \mathbf{J}(t) = -\frac{1}{\Omega} \operatorname{Im}\left[\sum_{n\kk}f_{n\kk}\langle\psi_{n\kk}|\boldsymbol{\nabla}|\psi_{n\kk}\rangle\right]
                    -\frac{N\mathbf{A}(t)}{\Omega}.
 \end{equation}
The expression that is actually implemented is a little different since we work in the 
PAW approach and due to some subtleties related to gauge invariance and non-local 
potentials~\cite{Pemmaraju2018,Sato2021}.
Finally, the optical conductivity can be obtained from Ohm's law in frequency space using the 
Fourier transform of $\mathbf{J}(t)$ and the macroscopic dielectric function can easily be calculated 
from the conductivity tensor.
Further information can be obtained from the documentation and the associated tutorial 
available on the \abinit website~\cite{abinit_website} as well as in 
Refs.~\citenum{Pemmaraju2018,RodriguesPela2021,Sato2021}.

\subsection{\texorpdfstring{$GW$}{GW} one-body reduced density-matrix}
\label{subsect:excitations_RDM}
The $GW$ approximation is usually employed to improve the description of band gaps of materials~\cite{Hedin1965}.
Nevertheless, the Green's function provides access to the density-matrix ({\em i.e.}, $n_1({\rr}, {\rr}')=-\textrm{i}G({\rr}, {\rr}', 0^-)$), where $0^-$ refers to the time limit when $t-t^\prime$ goes to zero from below.
Therefore, it yields all the one-electron ground-state properties, such as the electronic density, the dipole moments, etc.
The fully correlated self-consistent Green's function ($G^{scGW}$) can be computed by solving the Hedin equations iteratively. Surprisingly, a cost-effective low-order approximation to $G^{scGW}$ can be obtained by using the linearized version of the Dyson equation~\cite{bruneval2021improved},
\begin{equation}
    G^{scGW} \approx G^{lin} = G_0 + G_0 \Sigma G_0,
\end{equation}
where only the non-interacting Green's function $G_0$ and the approximate self-energy ($\Sigma = iG_0 W_0$) are required. 

Among the new capabilities of \abinit, we have the computation of $G^{lin}$ and its associated density-matrix ($n_1^{lin}$). These objects are built using imaginary frequencies to define the screening and the self-energy ($\Sigma$). The calculation of these objects is triggered by using \abivar{gwcalctyp} 21 and \abivar{gw1rdm} 1 in \abivar{optdriver} 4. (Notice that the number of imaginary frequencies has to be provided with \abivar{nomegasi}. For typical calculations, the number of frequencies ranges from 20 to 100.) For building these objects, it is highly recommended to set the number of bands to be used in the construction of the self-energy (\abivar{nband}) equal to the number of bands that are updated (\abivar{bdgw}) for all {\bf k}-points used. These settings will ensure that the number of electrons per unit cell is preserved ($N_e$). In our implementation, $n_1^{lin}$ is always transformed to the natural orbital representation before printing and computing properties ({\em i.e.}, $n_1({\bf r},{\bf r}')=\sum_{\bf k} \sum_p n_{{\bf k}p} \phi_{{\bf k}p}({\bf r})\phi_{{\bf k}p}^*({\bf r}')$ where $n_{_{{\bf k}p}}$ are the occupation numbers and $\phi_{{\bf k}p}$ are the natural orbitals). The construction of these objects can be time-consuming and, in certain computing facilities, it might be required to relaunch the calculation to complete the job. Therefore, it is recommended to save the data produced using checkpoint files that are printed using \abivar{prtchkprdm} 1; then, these files can be later read by \abinit{} by setting \abivar{irdchkprdm} 1. Finally, as shown in Refs.~\citenum{bruneval2019improved,denawi2023gw}, we recommend building $G_0$ using the wavefunctions and energies obtained with a hybrid density functional approximation ({\em e.g.}, PBEh($\alpha$) with a large amount of exact exchange
$\alpha$, for example $\alpha \geq 0.75$).

\begin{figure}
    \centering
    \includegraphics[width=\columnwidth]{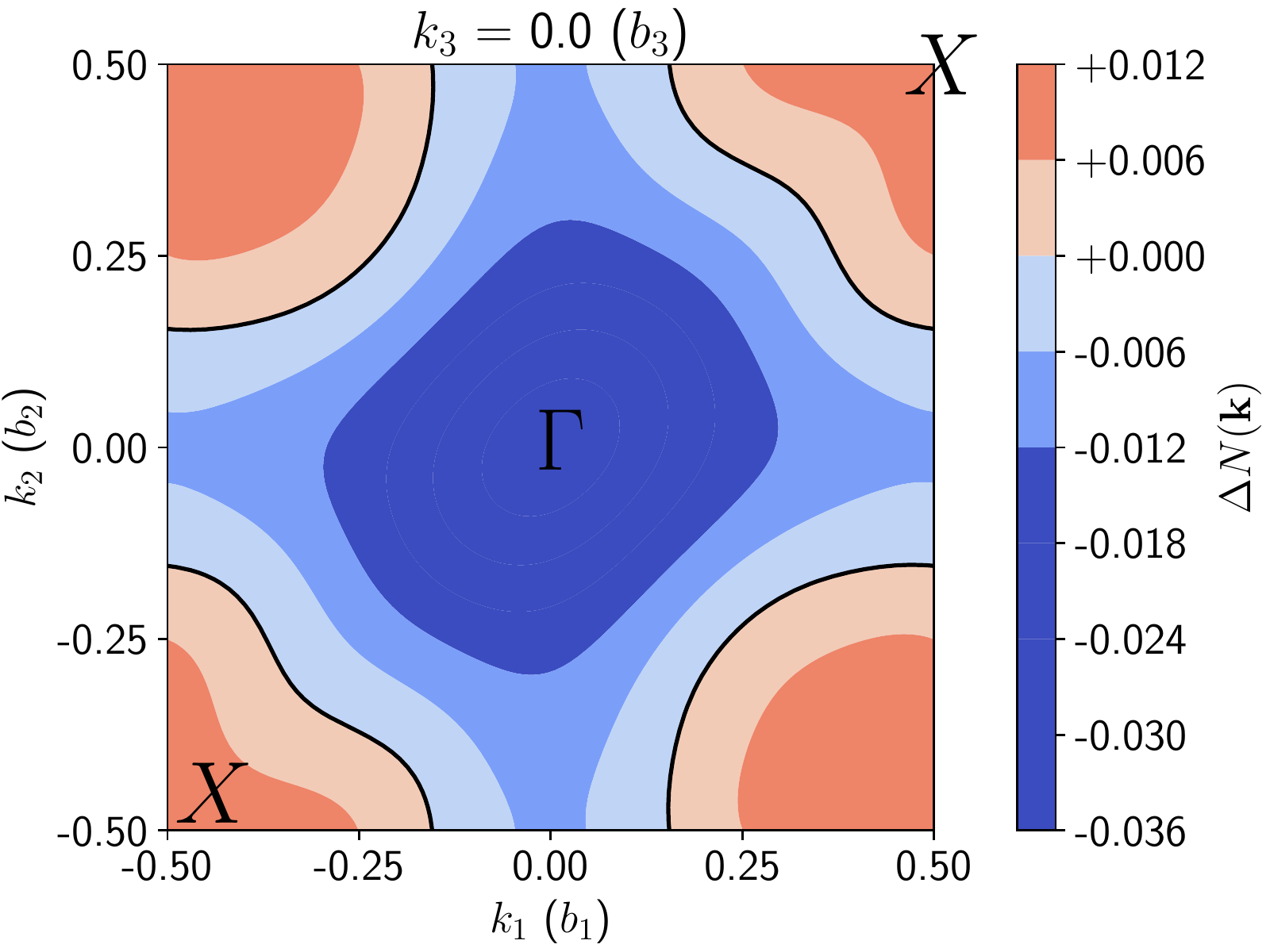}
    \caption{Electron count in zinc blende boron nitride as a function of $\kk$-point in the first Brillouin zone as obtained from the sum of the occupation numbers of $n_1^{lin}$, the linearized $GW$ density-matrix. For instance, a sizable electron transfer from $\Gamma$ to $X$ can be observed.}
    \label{fig:bn_nk_map}
\end{figure}

Once built, the density-matrix $n_1^{lin}$ can be used to evaluate the total energy 
\begin{equation}
 E=T[n_1^{lin}]+V_{ne}[n_1^{lin}]+E_\textrm{H}[n_1^{lin}]+E_x[n_1^{lin}]+ V_{nn}+E_c,   
\end{equation}
where the correlation energy ($E_c$) can be approximated by the Galitskii-Migdal correlation energy. The latter can now be computed at a vanishing cost during the construction of the screening ({\em i.e.}, in \abivar{optdriver} 3) by setting \abivar{gwgmcorr} 1. Total energies can be used to search for equilibrium geometries as shown in Ref.~\citenum{denawi2023gw}.
Finally, let us highlight a less known effect that is caused by electronic correlation: a rearrangement of the electronic density among the ${\bf k}$-points, {\em i.e.}, $N({\bf k})=\sum_p n_{{\bf k}p}\neq N_e$ and $N_e=\sum_{\bf k} N({\bf k})$. Interestingly, we can plot these rearrangements as $\Delta N({\bf k})=N({\bf k})- N_e$ in the whole Brillouin zone thanks to an \abipy{} utility that can be used to perform the interpolation of any ${\bf k}$-point-dependent obtained on the finite grid employed in the calculation (see Fig.~\ref{fig:bn_nk_map} that shows zinc blende boron nitride).
To obtain examples of the usage of all these new capabilities, the user is referred to the  \abivar{v9/t33--t37.abi} input files in the test suite.

\subsection{Low-scaling \texorpdfstring{$GW$}{GW} and RPA}

Since version 3.1, released about 25 years ago, \abinit{} has provided the possibility to compute quasi-particle (QP) corrections 
at the \textit{GW} level using a formalism in the frequency and Fourier domain, in which 
the irreducible polarizability $\chi$ and the self-energy $\Sigma$ are expressed in terms of a summation over bands~\cite{Onida2002, Aulbur2001, Golze2019}.
This traditional \textit{GW} algorithm scales quartically with the number of atoms in the unit cell 
and quadratically with the number of $\kk$-points, although the prefactor can be reduced by exploiting symmetries.
In recent times, however, there has been a renewed interest in formulating Hedin’s equations 
following the seminal work of Rojas and collaborators~\cite{Rojas1995, Rieger1999, Steinbeck2000},
in which the computation of $\chi$ and $\Sigma$ is performed in real space and imaginary time,
which allows cubic scaling with respect to system size and linear scaling with the number 
of $\kk$-points~\cite{Kaltak2014, Liu2016}. 
In the case of plane-wave-based codes, one relies on fast Fourier transforms (FFTs) to convert 
functions between real space and reciprocal space.
For instance, the equation for $W$, which is a convolution 
between $\epsilon^{-1}$ and the bare Coulomb interaction, 
is best solved in Fourier space and frequency domain.

One challenge in applying this approach lies in the substantial memory 
and storage costs required to store the $\chi$ 
and the screened interaction $W$, especially if one uses homogeneous meshes to sample the imaginary axis, 
as done in Rojas' original approach~\cite{Rojas1995}.
To reduce the computational effort, Kaltak \textit{et al.} proposed to use non-uniform imaginary-time 
and imaginary-frequency grids (\textsc{minimax grids}) in conjunction with inhomogeneous sine/cosine transforms 
to effectively reduce the memory cost~\cite{Kaltak2014, Liu2016}.
Extensive validation tests revealed that 20 minimax points are usually sufficient to achieve 
a difference below 10 meV between the \textit{GWR} gaps and the ones computed 
with a more expensive full-frequency integration of the self-energy~\cite{Liu2016}, 
and similar conclusions have been reported in a recent work using the \abinit implementation~\cite{Azizi2024}.
To activate the GWR driver in \abinit, one has to use \abivar{optdriver} = 6 
and choose the desired computation via \abivar{gwr\_task}. 
Various options are available, including direct diagonalization of the KS Hamiltonian using Scalapack, 
one-shot GW, and self-consistent GW calculations for energies and total energies at the RPA level.
Energy-only self-consistent calculations with GWR are notably simpler than in the conventional GW code, 
as there is no need to chain multiple screening/sigma calculations. 
The self-consistent loop and stopping criterion are directly implemented in Fortran, streamlining the process. 
A new tutorial explaining how to use GWR for a one-shot calculation and perform convergence studies has been added to the official documentation.

\subsection{Dynamical mean-field theory}
\label{subsect:excitations_dmft}
The combination of DFT with dynamical mean-field theory (DMFT)~\cite{RevModPhys.68.13} has been fruitful to describe strongly correlated systems.
Here, we review some of the recent implementations related to the DFT+DMFT implementations and calculation of effective interactions in \abinit, which were last reviewed in Refs.~\citenum{Romero2020,Gonze2020}.

\begin{figure}[!ht]
    \centering
    \includegraphics[width=1.0\linewidth,clip]{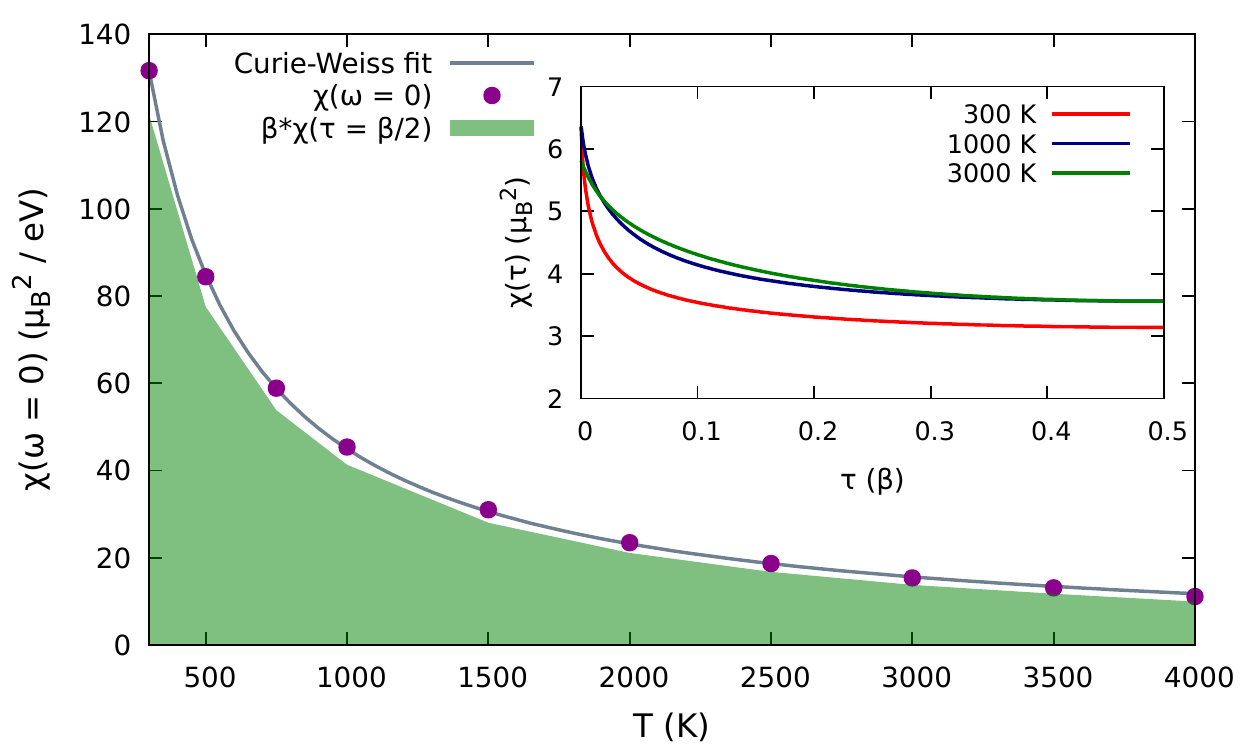}
    \caption{Temperature dependence of the static local magnetic susceptibility $\chi(\omega = 0)$ calculated for iron paramagnetic bcc (purple bullet)\cite{Gendron2022}. The grey line corresponds to a fit of the data in the Curie-Weiss regime, whereas the green area ($\beta \chi(\tau=\beta/2)$) is the contribution to the static susceptibility arising from the persisting local moment. The inset presents the corresponding local magnetic susceptibilities in imaginary time $\chi(\tau)$ for selected temperatures.}
    \label{fig:CurieIron}
\end{figure}

Firstly, the DFT+DMFT loop ({\abivar{usedmft} = 1)} has been optimized for efficiency. In particular, the calculation of the Green's function is now one order of magnitude faster, releasing a bottleneck for the study of super-cells.
Using the density-density internal continuous-time quantum Monte Carlo (CTQMC) solver (\abivar{dmft\_solv} = 5 or 8), the calculation of the configuration weights for the occupation of the correlated orbitals~\cite{Gendron2022} and local susceptibilities are now available using {\abivar{dmftctqmc\_localprop}}. 
The local spin-only, magnetic and charge susceptibilities are computed as $\chi^{\text{local}}(\tau)=\langle\hat{O}(\tau)\hat{O}(0)\rangle$ where $\hat{O}(\tau)$ is defined as $\hat{n}^{\uparrow}(\tau) - \hat{n}^{\downarrow}(\tau)=g_e\hat{S^z}(\tau)$, $(\hat{L}^{z}+g_e\hat{S}^{z})(\tau)$ and $\hat{n}(\tau) - \langle n \rangle$, respectively. Fig. \ref{fig:CurieIron} presents the Curie-Weiss behavior exhibited by the calculated local spin-only susceptibility in paramagnetic bcc iron.

Secondly, the C++ invocation (\abivar{dmft\_solv}=6, 7) of the CTMQC library {\textsc{triqs/cthyb}}~\cite{SETH2016274} to solve the DMFT impurity is extended to version v3.3 and will be available in the next version of \abinit{}.

Thirdly, the user can now construct the local orbitals using the interface to {\textsc{Wannier90}}~\cite{Pizzi_2020} and use the python invocation to delegate the Weiss field construction, impurity solving and extraction of the new density (\abivar{usedmft} = 10), using for example {\textsc{triqs/dfttools}}~\cite{AICHHORN2016200}.

\subsection{First-principles U and J parameters}
\label{subsect:LRUJ}
The \abinit code calculates two effective interactions used in DFT+U and DMFT, the Hubbard U and Hund's coupling J parameters. Their implementations have been strongly refactored and improved in the latest version. Firstly, the constrained random phase approximation (cRPA) has been generalized in order to compute interactions involving two different orbitals on the same atom or two different atoms~\cite{Moree2021,Outerovitch2023}. 

\begin{figure}[th]
    \centering
    \includegraphics[width=\columnwidth,trim={0cm 0cm 1.5cm 0cm},clip]{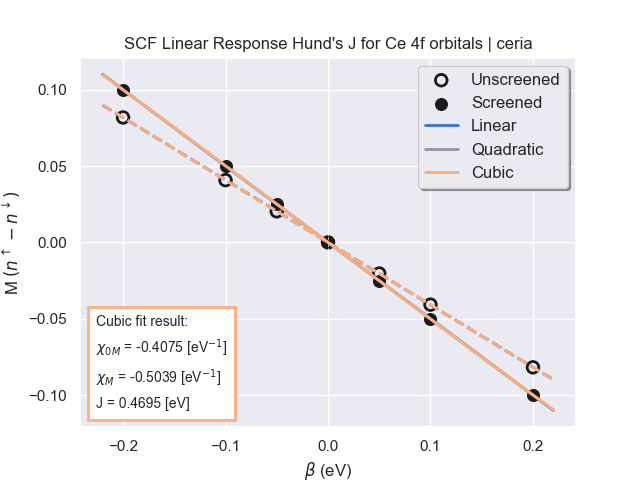}
    \caption{AbiPy plot demonstrating the spin-population response with respect to $\beta$ perturbation, used to calculate Hund's J for Ce $4f$ orbitals in CeO$_2$. Polynomial regression and uncertainty analysis is computed with the \texttt{lrUJ} utility.}
    \label{lrUJ_AbiPy}
\end{figure}

Secondly, the SCF linear response (LR) facilities 
in \abinit have been renovated and extended.  Automated calculations of the scalar Hubbard U~\cite{cococcioni2005} and 
now also Hund's J~\cite{Linscott2018}
are available, for use with
DFT+U, various DFT+U+J functionals, and potentially also with DMFT.
Spin-dependent U parameters may be computed or, 
for non-spin-polarized systems, 
it is possible
to simultaneously compute U and J with a single set
of finite-differences calculations~\cite{Lambert2023}.
The \texttt{lrUJ} post-processing utility is designed to handle the non-linear response
that may arise in some systems. It provides streamlined regression, error analysis, and visualization of output LR data
through its interface with AbiPy plotting.
For a detailed description
of the \texttt{lrUJ} functionality
for automated U and J calculation, see Ref.~\citenum{MacEnulty2024}; 
for a user's guide, see Ref.~\citenum{lruj_tutorial}. 
A recent application to NiO 
assesses the Hubbard projector 
and spin-state dependence of the
parameters, 
see Ref.~\citenum{MacEnulty2023}. 
Figure \ref{lrUJ_AbiPy} shows a sample SCF linear response calculation for scalar J $=-\left[ \chi_{0 M}^{-1}-\chi^{-1}_M \right]$ in CeO$_2$. Here $\chi_M$ ($\chi_{0 M}$) is the  screened (unscreened) response of the $4f$ subspace spin-population $M$, with respect to a Zeeman type spin perturbation of strength $\pm \beta$, which is uniform in the subspace, 
following Ref.~\cite{Linscott2018,Lambert2023}.

\subsection{Interface to Cc4s, a coupled-cluster package for solids}
\label{subsect:excitations_cc4s}

Cc4s (Coupled Cluster for Solids) is an open-source software package~\cite{cc4s_website}
for highly accurate post-Hartree-Fock studies of electronic properties in periodic systems.
It implements CCSD (coupled cluster with single and double excitations),
and also perturbative triple excitations (CCSD(T)) to reach higher accuracy.
Cc4s utilizes advanced high-performance libraries like
the cyclops tensor framework (CTF)~\cite{Solomonik2014}, 
an efficient parallel tensor contraction engine, 
and computational paradigms such as MPI
for parallelization.

In a typical calculation, the ground-state part of \abinit{} is used to perform a Hartree-Fock calculation including empty states
for the sole $\Gamma$-point.
At this point, one uses \abivar{optdriver} 6 and \abivar{gwr\_task} \texttt{"CC4S\_FROM\_WFK"} to activate the interface between the two codes.
\abinit{} reads the single-particle wavefunctions from
\abivar{getwfk\_filepath} and then computes the Coulomb vertex defined by
\begin{equation}
    \Gamma^{i \GG}_{j} =
    \sqrt{\frac{4 \pi}{\GG^2}}
    \int 
    e^{-i\GG \cdot\rr}
    \varphi^{*}_i(\rr)
    \varphi_{j}(\rr)
    \ \mathrm{d} \rr
    ,
\end{equation}
%
%
where $i$ and $j$ are state indices and where $\GG$ runs over the plane waves defined by the cutoff energy \abivar{ecuteps}.
The delicate treatment of the divergence when $\GG=0$ is governed by input variable \abivar{gw\_icutcoul}.
Note that \abinit can take advantage of the time-reversal symmetry at $\Gamma$
and the real-valued nature of the wavefunctions so to reduce computational burden and file sizes by a factor 2.

Cc4s is independent of the nature of the auxiliary basis that is used to expand the Coulomb integrals.
In the present case, the auxiliary basis is composed of plane waves.
Cc4s obtains the two-electron integrals $( i j | k l )$ through a decomposition
using the Coulomb vertex $\Gamma^{i \GG}_{j}$: 
\begin{equation}
  ( i j | k l ) = \sum_{\GG} {\Gamma^{i \GG}_{j}} {\Gamma^{l \GG}_{k}}^* .
\end{equation}
The computation of the matrix elements $\Gamma^{i \GG}_{j}$ is parallelized using MPI and operates on MPI-distributed wavefunctions.
Both NC and PAW pseudopotentials are supported.
YAML files containing dimensions and metadata are automatically generated by the interface.
After this step, all the necessary files for Cc4s are available.
Examples are available in the gwr\_suite directory of the test suite.

%% file: 5_HPC.tex
\section{Numerical Optimization and High Performance Computing }
\label{sect:HPC}

Considerable effort over the last few years has gone into improving
\abinit performance.
As more and more supercomputer architectures rely on GPU accelerators, it is now very important to be able to use these architectures efficiently.
Section~\ref{subsect:HPC_GPU} presents in detail a significant improvement of GPU implementations for ground state and DFPT computations.
A new implementation of non-local operations in ground-state computations is presented in Section~\ref{subsect:HPC_nonloc}.
The parallelization of bands in DFPT calculations is presented in Section~\ref{subsect:HPC_band_paral}, which
allows for large DFPT calculations with minimal memory usage.
Finally, ground-state computations can also be accelerated with the Residual Minimization Method, Direct Inversion in the Iterative Subspace (RMM-DIIS), as described in Section \ref{subsect:HPC_rmm_diis}.

\subsection{GPU use for the ground state and DFPT} 
\label{subsect:HPC_GPU}

A major milestone in the development of \abinit was its successful adaptation to GPU accelerators. An initial version of \abinit was ported to GPUs in 
2013~\cite{Gonze2016}, but it had become obsolete, did not meet current coding standards, and exhibited several limitations. In that initial version,
\abinit could only run on \textsc{nvidia} accelerators because the code was directly written in the \textsc{cuda} language. The performance achieved aligned with decade-old standards, yielding a speed-up of 3 to 5, as measured by a straightforward metric comparing execution on $N$ CPU cores to execution on $N$ CPU cores augmented by $N$ GPU accelerators.

The version of \abinit presented here offers a much more modern and efficient GPU implementation. The GPU code has been rewritten from scratch, completely rethinking the programming model and the iterative diagonalization algorithms.

Among the various possibilities, the programming model selected here relies on \texttt{OpenMP offload} directives (available since \texttt{openMP} version 4.0, with full support in version 5.0)~\cite{openmp_website}. The \texttt{OpenMP} standard appears to be a sustainable choice, because it is increasingly supported by many compilers and allows for the addressing of different types of accelerators. It is worth noting that a limited version of \abinit using the \texttt{Kokkos}~\cite{kokkos_website} performance library is also available in accelerated mode, giving access only to the total energy of the ground state.

Programming with \texttt{OpenMP} directives offers numerous advantages. It is minimally verbose and intrusive in the code. Only a minimal set of directives is required for the porting process: \texttt{OMP TARGET TEAMS DISTRIBUTE} to distribute loops across threads, \texttt{OMP TARGET UPDATE TO/FROM} to perform transfers between CPUs and GPUs, and \texttt{OMP TARGET ENTER/EXIT DATA MAP} to allocate memory on the GPU and associate it with CPU memory. These latter directives are optional when using modern features related to managed \textit{unified memory}. To ensure safe coding practices, all directives related to the GPU in \abinit are safeguarded and rendered visible to the compiler exclusively when GPU compilation is enabled.

The strategy for porting \abinit to GPUs is based on the following ideas:
\begin{itemize}
\item Intensively use "batch processing" mode to perform multiple operations in parallel by replacing loop indices with an additional dimension in arrays. For instance, executing multiple concurrent Fourier transforms using an array of wave functions is more efficient on GPUs than performing a loop of Fourier transforms on individual wave functions.
\item  Minimize host-device and device-host transfers. Although significant progress has been made in these operations, they remain costly.
\item  Use, whenever possible, the libraries provided by manufacturers for elementary operations such as linear algebra, matrix algebra, or Fourier transforms.
\item  Prioritize iterative diagonalization algorithms that apply the Hamiltonian to a vector over those based on subspace diagonalization procedures. This results in a preference for diagonalization algorithms that employ spectrum filtering over those that rely on block minimization~\cite{Levitt2015}. More details on this follow.
\end{itemize}

The code is fully operational on \textsc{nvidia} and \textsc{amd} accelerators. However, achieving the expected speed-up requires accelerators with strong double-precision performance. As mentioned earlier, performance heavily relies on the libraries provided by manufacturers (\texttt{[cu/roc]BLAS}~\cite{cublas_website,rocblas_website}, \texttt{[cu/roc]FFT~\cite{cufft_website,rocfft_website}}, \texttt{[cu/roc]Solver~\cite{cusolver_website,rocsolver_website}}). It is also possible to activate, when available, the \textit{GPU-aware} mode of the \texttt{MPI} library. The performance gain is then substantial. Tests have also been conclusive on NVIDIA using \textit{GPUDirect} communications.

To compile the GPU-compatible code, specific options need to be enabled in the \textit{build system}. Detailed instructions and example configuration files for common architectures are provided in the official \abinit documentation. The GPU code can be generated using either the \texttt{Autotools} build system~\cite{autotools_website} or the new \texttt{CMake} build system~\cite{cmake_website}.

To enable the execution of \abinit on GPU, simply add the keyword \abivar{gpu\_option} in the input file.

All \abinit functionalities related to ground state calculations are available in the GPU-accelerated version. This includes the calculation of energy, forces, and stresses for all types of magnetism, including non-collinear magnetism, Hubbard correction in DFT+U, and spin-orbit coupling. All types of exchange and correlation functionals are accessible to the GPU code, including meta-GGA functionals and hybrid functionals. For the latter, the speed-up is notably significant. The calculation of the Green's function used in the Dynamical Mean Field Theory (DMFT) method has also been ported to GPU accelerators.

Most functionalities related to linear response calculations within the Density-Functional Perturbation Theory (DFPT) framework can be executed on GPU accelerators.
This covers the calculation of vibrational spectra, responses to electric fields, and the elastic tensor.
However, the speed-up for these calculations is not as strong as for ground state calculations, as the code structure is less suited to intensive batch processing.
Non-linear response calculations are not covered here.

For ground state and response function calculations, all functionalities are available within the framework of norm-conserving pseudopotentials and the Projector Augmented-Wave (PAW) approach.
\begin{figure*}
  \centering
  \includegraphics[scale=0.5]{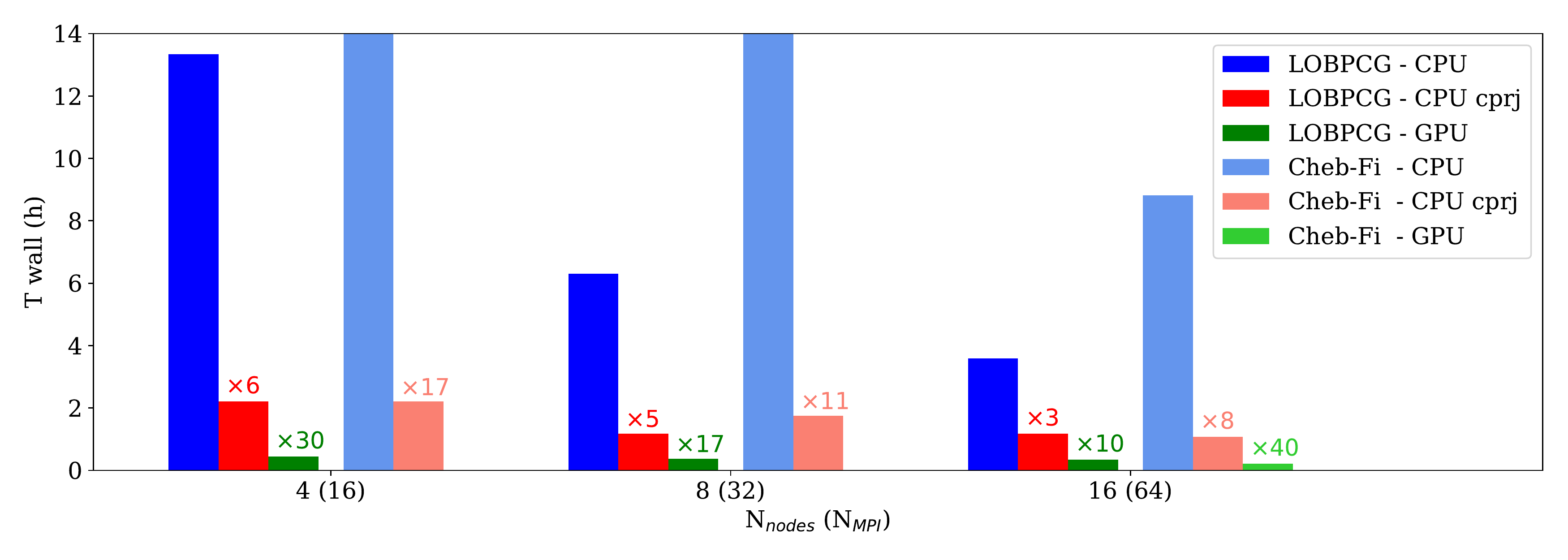}
  \cprotect\caption{Execution time of \abinit for determining the ground state energy, 
  forces and the stress tensor of a 1,280 atoms crystal of \ce{Ga2O3} with 6,144 electronic bands, PAW pseudopotentials and a cutoff energy of 18~Ha.
  This study is performed on \textit{Topaze} supercomputer at the TGCC computing center~\cite{tgcc_website}, Bruyères-le-Châtel, France.
  Each compute node consists of 2 AMD Milan EPYC 7763 chips (64 cores, totaling 128 cores per node),
  256 GB of memory per node, and 4 Nvidia A100 80GB GPU accelerators.
  The calculation is first performed on CPU cores alone using the default implementation (blue histograms),
  then using \abivar{cprj\_in\_memory~1} as explained in section \ref{subsect:HPC_nonloc} (red histograms),
  then using GPU accelerators with \abivar{gpu\_option~"GPU\_OPENMP"} as explained in section \ref{subsect:HPC_GPU} (green histograms).
  Histograms on the left(right)-hand side show results obtained with LOBPCG (Chebyshev Filtering) algorithms.
  The speed-ups with respect to corresponding blue histograms are indicated.
  All computations use 4 MPI processes per node, so CPU (GPU) computations are done with 32 OpenMP threads (one GPU) per MPI process.
  LOBPCG parameters are \abivar{nblock\_lobpcg~4} and \abivar{nline~4} and Chebyshev Filtering runs are done with \abivar{mdeg\_filter~10}.
  All runs converge with 14 electronic iterations.
  Computations with Chebyshev Filtering on GPU using 4 nodes or 8 nodes could not be done due to memory requirements.
  \abinit is compiled with Intel+MKL for CPU and nvhpc+CuBLAS+CuFFT for GPU.}
  \label{fig:cpu_gpu_speedup}
\end{figure*}
The use of GPU accelerators is compatible with all other levels of parallelism in the code.
However, since most of the parallelism is handled by the graphics cards, it is necessary to reduce the number of \texttt{MPI} processes or \texttt{OpenMP} threads on the CPU.
Using GPUs requires users to adjust their habits and configure their calculations to utilize significantly fewer compute nodes.

The iterative algorithm designed to diagonalize the Hamiltonian in a plane-wave basis required adaptation.
In a plane-wave DFT code, this algorithm, designated as ``matrix-free'', relies on two primary operations: first, applying the Hamiltonian to a set of vectors, and secondly, diagonalizing it within the subspace of eigenvectors (known as the Rayleigh-Ritz procedure).
The application of the Hamiltonian is well-suited for GPUs, benefiting from efficient batch processing.
Conversely, the subspace diagonalization necessitates communication between GPUs and is less efficient.
Therefore, it is crucial to minimize calls to the Rayleigh-Ritz procedure.

For these reasons, it is recommended to employ an algorithm based on spectral filtering, such as the \textit{Chebyshev filtering} algorithm implemented in \abinit since 2015~\cite{Levitt2015}.
This approach minimizes calls to the Rayleigh-Ritz procedure and intensively utilizes the Hamiltonian.
Given the high efficiency of applying the Hamiltonian on GPUs, it is feasible to increase the degree of the filtering polynomial without significantly compromising performance, while notably enhancing the quality of the eigenvectors.

An example of the performance achieved with \abinit using GPU accelerators is presented in Figure~\ref{fig:cpu_gpu_speedup}.
This example involves determining 6,144 electronic bands for gallium oxide through a \textit{strong scaling} study.
It is important to note that the metric used is now a "node-to-node" comparison, which differs significantly from the 2013 version.
Using this comparison criterion, a speed-up of 10 to 30 is observed when GPU accelerators are activated for LOBPCG (Locally Optimal Block Preconditioned Conjugate Gradient)~\cite{Bottin2008}, and a speed-up of 40 for Chebyshev Filtering algorithm.
The speed-up naturally depends on the computational load, so lower speed-up could be obtained for smaller systems.

Upon closely examining the performance, it appears that the GPU code exhibits lower scalability compared to the CPU code, aligning with \textit{Amdahl's law}.
Increasing the number of GPU nodes proves less efficient than performing the same calculation on CPUs.
There is no need to increase the number of GPUs, as is typically done with CPUs.
The execution time on 4 GPU nodes is faster than on 16 CPU nodes, leading to significant energy savings.
This benefit is particularly noteworthy but can only be realized if \abinit users adjust the resources requested for their calculations accordingly.

\subsection{Optimization of non-local operations on CPU}
\label{subsect:HPC_nonloc}

The non-local part of the Hamiltonian $H^{\text{nl}}$ applied to a wavefunction $\psi_n$ writes :
\begin{equation}\label{eq:ham-nl}
H^{\text{nl}}_{\kk}\ket{\psi_{n\kk}} = \sum_{a=1}^{N_{at}} \sum_{ij} \ket{p_{a,i,\kk}} h_{a,ij} \braket{p_{a,j,\kk}}{\psi_{n\kk}}
\end{equation}
where $N_{at}$ is the number of atoms and $i$ or $j$ stand for $n,l,m$ atomic quantum numbers.
$p_{a,i,\kk}$ are the projectors on atomic orbitals, obtained from the pseudopotentials.
For Norm-Conserving pseudopotentials, $h_{a,ij}$ are diagonal matrices whose coefficients are Kleynman-Bylander energies.
In the PAW method, $h_{a,ij}$ are obtained from several self-consistent terms.
We note that the non-local part of the overlap operator has the same structure.

Evaluating Eq.~\ref{eq:ham-nl} is the most costly part of the Hamiltonian for big systems.
Indeed, its computational complexity (when applied to one band) is proportional to $N_{at}^2$,
whereas the complexities of the kinetic part and the local part using fast Fourier transforms are $N_{at}$ and $N_{at} \ln(N_{at})$, respectively.

The first step is to compute the coefficients $c_{a,i,n,\kk}~=~\braket{p_{a,i,\kk}}{\psi_{n\kk}}$, called ``cprj'' in the \abinit sources.
In the native version of the code, these coefficients and the projector plane-wave components $p_{a,i,\kk}(\GG)$ are computed on-the-fly.

Now an alternative implementation is available, in which the coefficients $p_{a,i,\kk}(\GG)$ and $c_{a,i,n,\kk}$ (for all bands) are computed and stored in memory when solving the Kohn-Sham equations.
This way, all operations needed to compute non-local terms can be expressed in terms of matrix-matrix multiplications, which can be efficiently obtained with BLAS calls.
Using this framework, new versions of LOBPCG and Chebyshev filtering algorithms have been implemented to reduce the number of non-local operations, without any approximation.

This new implementation is activated with the input variable \abivar{cprj\_in\_memory~1}, as explained in the \abinit documentation.
Its usage is restricted to ground state computations done on CPUs.

This performance optimization follows, and fully takes advantage of, the work introduced in the section 3.8 of Ref.~\citenum{Gonze2020}.
In particular, new non-local routines and algorithms have been implemented using an abstract layer which allows an efficient use of MPI and OpenMP parallelisms.
The later is efficient only if \abinit is linked with a multi-threaded BLAS/LAPACK library.

Performances obtained with \abivar{cprj\_in\_memory~1} are presented in the Figure~\ref{fig:cpu_gpu_speedup} for a system with 1280 atoms, showing a speed-up of 3 to 6 for LOBPCG, and 8 to 17 for Chebyshev Filtering algorithm.

\subsection{Band parallelized DFPT} 
\label{subsect:HPC_band_paral}
Low energy response properties can be calculated within the framework of DFT using standard perturbation theory (DFPT), as described in Section~\ref{sect:DFPT}. This subsection details extensions to the parallel HPC implementation of DFPT, distributing bands and $\kk$-points between MPI threads on different processors, to reduce both memory footprint and computing time.
\abinit implements the variational Sternheimer equations~\cite{Sternheimer1954}
\begin{equation}
    P_c (H^{(0)}-\epsilon^{(0)})P_c \psi^{(1)} = - P_c H^{(1)} \psi^{(0)}
\label{eq:sternheimer}
\end{equation}
with $^{(0)}$ and $^{(1)}$ denoting ground state and first order perturbed quantities, and $P_c = 1 - \sum_j^{occ} |\psi_j^{(0)}\rangle\langle \psi_j^{(0)}|$ is the projector on the unoccupied subspace of bands. The mixed derivatives ($j_1, j_2$) of the total energy are obtained using non-variational expressions (see Ref.~\citenum{Gonze1997} for details and notations) summing over occupied bands for each perturbed $\psi^{j_2}$.
Because of the particular orthogonalization conditions (with respect to the unperturbed wave functions) and choice of gauge in DFPT,
the equations are actually easier to solve numerically than the ground state ones. 

As for the ground state KS equations, Eq.~\ref{eq:sternheimer} is trivially parallel in the wave vector $\kk$, only requiring the wave functions $\psi^{(1)}_\kk$, $\psi^{(0)}_\kk$, and $\psi^{(0)}_{\kq}$, which has been exploited already in previous versions of \abinit to parallelize both memory and computation using Message Passing Interface (MPI) threads.

When one considers larger or more disordered systems, however, the numerical effort goes up in number of bands and/or plane waves, while the number of $\kk$-points is reduced. One can still distribute the calculation of each band $\psi_i^{(1)}$ on a different processor, but the $P_c$ impose orthogonalization to all of the $\psi_j^{(0)}$ individually, and thus a memory footprint per MPI thread which increases with the number of bands.

In the present version of \abinit this problem has been solved by distributing fully the bands of $\psi^{(1)}$ and $\psi^{(0)}$ in blocks on each MPI thread in a band communicator of size $\mathrm{NP}_{band}$, and introducing additional communication steps. This is necessary both in the Conjugate Gradient SCF cycle, applying $P_c$ and in evaluating Eq.~\ref{eq:sternheimer}. 
A global loop over all bands implies all processors act on a given band $i$. $\psi_i^{(1)}$ or $H^{(1)}\psi_i^{(0)}$ are MPI\_BCAST by the owner of band $i$ at the beginning of the loop, orthogonalized over all bands by each processor individually, and then collected (MPI\_SUM) and saved by its owner at the end of the loop. In order for this last step to stay additive while keeping the existing routines for projection, care must be taken to remove the $\mathrm{NP}_{band}$-fold double counting of the constant term in $P_c$.

The band block distribution and synchronous operation on each band impose that the number of bands is an integer multiple of the number of processors in the band communicator. A safe practice is to start with a large number of empty bands for $\psi^{(0)}$ (a cheap calculation) as the last bands can be neglected in the DFPT step if need be. With this method, large DFPT calculations up to $\sim 200$ atoms have been carried out with a minimal memory footprint.

\begin{figure}
  \centering
  \includegraphics[scale=0.8]{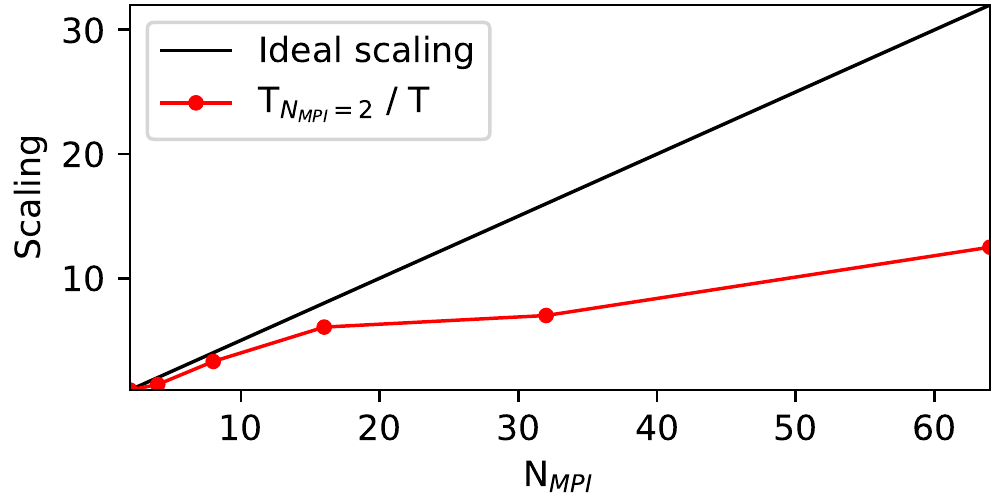}
  \cprotect\caption{Scaling of band parallelism for DFPT on CPUs. The CPU architecture and system are the same as in \ref{fig:cpu_gpu_speedup}, but with 80 atoms and 384 bands. The run with $N_{MPI}=1$ could not be completed due to memory requirements, so $N_{MPI}=2$ is used as reference.}
  \label{fig:dfpt_scaling}
\end{figure}

A scaling example is presented in Figure~\ref{fig:dfpt_scaling} for a system with 80 atoms and 384 bands. The overhead in communication is visible from 16 CPUs per band pool onwards, and should be balanced with the memory reduction needed on a specific architecture.

\subsection{RMM DIIS} 
\label{subsect:HPC_rmm_diis}

The Residual Minimization Method, Direct Inversion in the Iterative Subspace (RMM-DIIS)~\cite{Kresse1996},is an eigensolver designed to accelerate ground-state computations, structural relaxations, and molecular dynamics simulations. 
RMM-DIIS is compatible with both NC and PAW pseudopotentials and fully supports the \abivar{ paral\_kgb} parallelization scheme, 
though PAW calculations tend to be more sensitive to internal optimizations and may not achieve the same speedup as NC cases.

RMM-DIIS is typically used alongside another eigensolver ({\em e.g.}, CG or LOBPCG) for initial SCF iterations, 
as it does not guarantee convergence to the ground state but finds the eigenvector closest to the initial trial state. 
The accuracy and stability of RMM-DIIS depend heavily on the quality of these initial wavefunctions.
%
RMM-DIIS generally reduces wall-time per iteration since it avoids explicit orthogonalization, 
requiring only a single full-band Cholesky orthogonalization per SCF step. 
It can be up to twice as fast as CG/LOBPCG,
though its subspace rotation 
and Cholesky steps scale poorly with MPI and may dominate wall-time in large systems.
To activate the new algorithm  
it is sufficient to set {\abivar{rmm\_diis}} = 1.to activate the eigenvalue solver
after {\abivar{rmm\_diis}} SCF iterations.
%

Since RMM-DIIS is rather sensitive to the quality of the initial trial wavefunctions, \abinit implements the possibility of initializing the wavefunctions using a linear combination of (pseudized) atomic orbitals via \abivar{wfinit}.
At present, this option is only available for NC pseudos in the UPF2 format.

%% file: 6_multiscale.tex
\section{Multiscale and Second Principles Methods}
\label{sect:multiscale}

While density functional theory (DFT) constitutes an essential tool in modern materials science, its applicability is fundamentally limited by computational cost to small length scales (a few hundred atoms per cell) and timescales (a few picoseconds). These limitations make it impractical for exploring many pressing problems such as phase transitions, domain wall dynamics, and the complex interplay between structural, electronic, and spin degrees of freedom.
Tackling such challenges requires access to mesoscopic length scales, extended dynamical regimes, and realistic working conditions,
including finite temperatures, mechanical stresses, or applied electric fields without sacrificing the predictive power and accuracy that first-principles methods provide.
To overcome DFT limitations, second-principles and multiscale approaches have emerged as powerful strategies.
These methods construct effective models rooted in first-principles data, enabling simulations that retain the atomic resolution and preserve the quantum-mechanical accuracy of DFT while extending accessible spatial and temporal scales by orders of magnitude.
This section presents the evolution of the 
\multibinit and TDEP packages, which produce suitable effective approaches for mesoscale simulations of atom and/or spin dynamics.

\subsection{\multibinit {} second-principles models}
\label{subsect:multiscale_multibinit}

\multibinit is an integrated software package within the \abinit suite that enables the construction and application of second-principles effective models. It supports the entire workflow, {\em i.e.}, from the generation of first-principles data and the construction of the model to its use in large-scale simulations of materials properties under realistic operating conditions.
These second-principles models rely on parameters extracted from first-principles calculations, allowing simulations at much larger spatial and temporal scales and at a significantly lower computational cost, while
retaining the predictive power of ab initio methods. \multibinit currently treats lattice and spin degrees of freedom through three dedicated modules: (i) MB-EAP, for effective atomic potentials; (ii) MB-LWF, for lattice Wannier function-based effective Hamiltonians; and (iii) MB-SPIN, for spin Heisenberg Hamiltonians.

The software package has undergone methodological improvements, strengthening its capabilities in modeling
complex materials systems and improving both accuracy and computational efficiency.

\subsubsection{\multibinit-EAP}

%

\multibinit-EAP (MB-EAP) relies on the construction of effective atomic potentials to perform atomistic simulations of materials under working operating conditions such as finite temperature, electric fields or mechanical strain. 
Simulation capabilities include structural relaxations, molecular dynamics or hybrid Monte Carlo schemes for enhanced sampling of the potential energy surface.
Following Ref.~\citenum{Wojdel2013}, MB-EAP relies on a Taylor expansion of the Born-Oppenheimer energy surface around a stationary structure taken as a reference. The expansion is carried out in terms of individual atomic displacements and macroscopic strain degrees of freedom, with all coefficients determined from first-principles data. The energy is partitioned into harmonic and anharmonic contributions associated with atomic displacements,
macroscopic strains and their couplings, with the long-range dipole–dipole interaction treated explicitly. At the harmonic level, the coefficients correspond exactly to those from density functional perturbation theory and the input is the \abinit DDB file. At the anharmonic level, the most relevant terms are selected and their coefficients are fit to reproduce DFT energies, forces and stresses for a representative set of configurations
that properly sample the Born-Oppenheimer energy surface, while guaranteeing the boundedness of the model. A comprehensive overview of the methodology will be provided separately in Ref.~\citenum{Bastogne2025a}.

Recent developments have extended previous capabilities. Optimization of the generation and fitting of symmetry-adapted terms (SAT) ensures more accurate incorporation of symmetries. A selective SAT generation mechanism using number-of-body filtering enables computationally efficient inclusion of higher-order terms. Stress calculations include improved force-derived corrections. Algorithm optimizations make the software more efficient for large-scale simulations. The weight of some training set configurations can be adjusted for better accuracy in low energy ranges. 
The bounding terms now use the same body-count as the term to be bounded, 
promoting model simplicity and reducing the energy penalty introduced.

\begin{figure}[t!]
	\centering
	\includegraphics[width=0.99\linewidth]{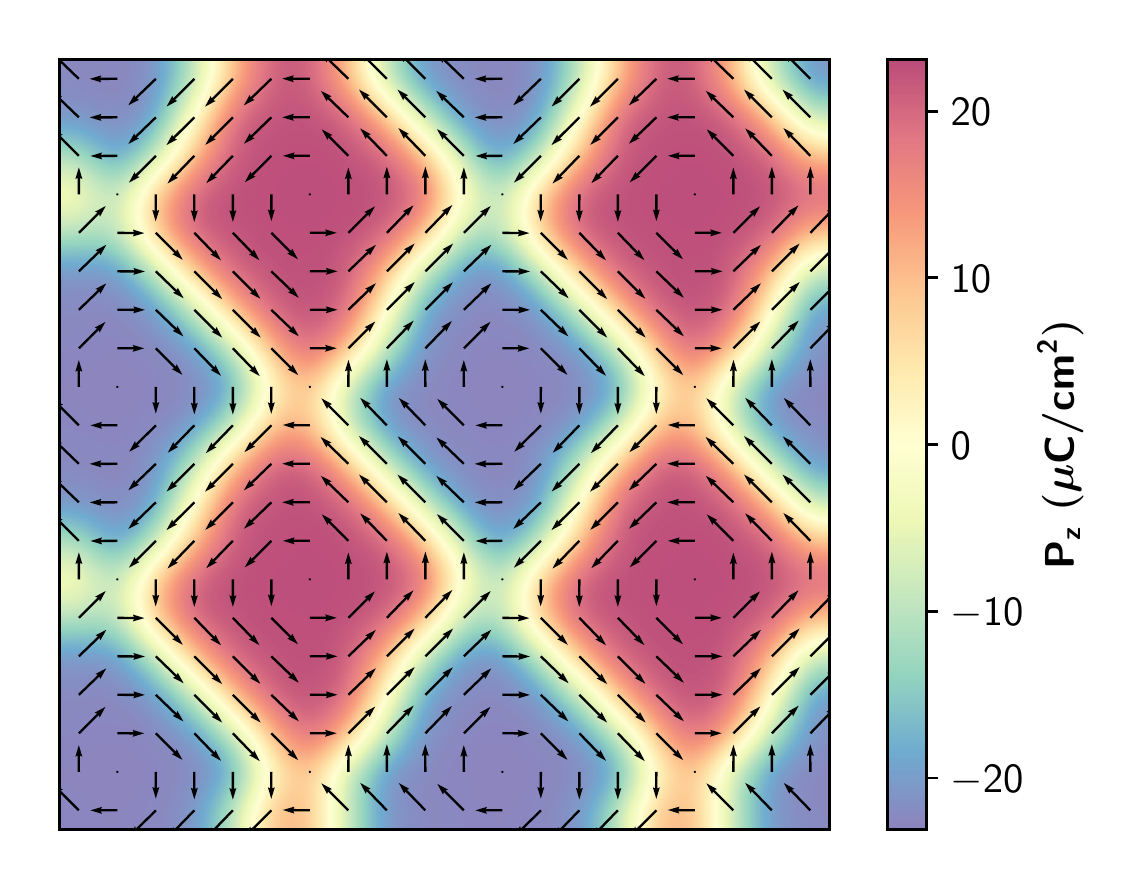}
    \caption[Multibinit]{Meron-antimeron polar texture stabilized in \ce{BaTiO3} by an acoustic 
    phonon excitation using Multibinit. Arrows show the amplitude of local electrical polarization in the plane, while the color bar shows the out-of-plane component. Adapted with permission from Ref.~\cite{Bastogne2024a}
    American Chemical Society}
    \label{fig:MB_idea}
\end{figure}

A totally new functionality concerns the possibility of performing simulations under spatially inhomogeneous and time-dependent external electric fields~\cite{Gomez2025a}. The field couples to the atoms through their Born effective charges, enabling the investigation of field-induced phenomena in polar materials. 
This feature opens the door to a broad range of applications, including the study of nonlinear and frequency-dependent dielectric responses, polarization switching under ultrafast or spatially patterned fields,
domain wall dynamics, and electrocaloric effects. It also enables the modeling of realistic device operating conditions, such as AC driving or field gradients.
An example is shown in Fig.~\ref{fig:MB_idea} with a polar texture generated by an excited phononic state.

MB-EAP is a versatile approach that can be applied to a broad range of materials systems. Recently, it has been used to develop accurate models for various perovskite oxides, including \ce{BaTiO3}~\cite{Bastogne2024a}, 
\ce{PbTiO3}~\cite{Bastogne2025b}, \ce{SrTiO3}~\cite{Bastogne2025a}, and 
\ce{CaTiO3}~\cite{Zhang2025}, with the models and related validation passports made freely accessible~\cite{Sriram-25}. 
The method proved successful in simulating structural phase transitions in temperature~\cite{Zhang2023a}, characterizing complex domain wall structures ~\cite{Alexander2025} or uncovering the existence of unexpected topological polar textures in ferroelectrics~\cite{Gomez2025b}. Available bulk models can also be combined to study superlattices or heterostructures~\cite{Zatterin2024}. Relying on physically transparent parameters, MB-EAP is also particularly relevant to discuss and rationalize the behavior of distinct compounds~\cite{Zhang2023b}. It can also be used to provide inputs to TDEP method (see below) at low computational cost~\cite{Bastogne2025b}. 

\subsubsection{\multibinit-LWF}

Lattice Wannier functions (LWFs) and related effective Hamiltonians are a powerful approach for developing minimal effective models for lattice dynamics.
They were pioneered by Zhong, Vanderbilt, and Rabe~\cite{rabe_1995_localized} in the 1990s. LWFs serve as localized basis sets in real space, analogous to electronic Wannier functions but specifically tailored for phonons. While individual atomic displacements provide a natural basis for structural distortions, LWFs can provide a more suitable representation for projecting within a restricted minimal subspace, when studying compounds exhibiting displacive structural phase transitions dominated by specific modes. 

As for \multibinit-EAP, the model is based on a Taylor expansion of the energy around a reference structure and in terms of selected degrees of freedom. The harmonic part is directly constructed by mapping from the harmonic interatomic force constants, and the anharmonic interaction parameters are fit to first-principles data. 

The key advantage of LWFs emerges in modeling anharmonic interactions, where computational complexity scales with the basis set size. By providing a minimal basis that captures essential physics, LWFs enable efficient and interpretable models for studying structural phase transitions and phonon-mediated phenomena. The reduced number of parameters required for fitting makes the approach computationally efficient, while the close relationship between LWFs and physical distortion patterns offers direct insight into underlying mechanisms. This approach complements MB-EAP and machine learning potentials, offering enhanced interpretability and fewer parameters.

The LaWaF package~\cite{he_2025_lawaf} has been developed to address key challenges in LWF construction, such as phonon band entanglement and long-range dipole-dipole interactions. The software constructs LWFs and their Hamiltonians, including both harmonic and anharmonic terms, outputting directly to \multibinit-compatible formats.

The \multibinit-LWF module implements both Metropolis Monte Carlo sampling and velocity-scaled lattice dynamics~\cite{berendsen_1984_molecular} which is similar to the usual molecular dynamics, but taking the amplitude of the LWF's as the degrees of freedom.
These capabilities have enabled investigations of structural-phase-transition-related phenomena in materials such as infinite-layered nickelates  
\ce{{\em R}NiO2}~\cite{zhang_2023_displacive,zhang_2023_rareearth} and 
\ce{LaAlO3}~\cite{jia_2022_dynamic}.

\subsubsection{\multibinit-SPIN}

While the spin dynamics implementation in \multibinit has maintained its stability with only bug fixes and minor improvements, significant advancements have been made in TB2J~\cite{he_2021_tb2j}, the related package for constructing the spin model parameters based on Heisenberg Hamiltonian which is used as the input of \multibinit-Spin, extending the use of \multibinit to a wider range of magnetic materials and phenomena. 

The TB2J package has undergone several updates to enhance its functionality and integration with \multibinit. Key improvements include the implementation of enhanced numerical methods for improved stability and computational efficiency, and the addition of magneto-crystalline anisotropy energy (MAE) calculation capabilities. The package's compatibility has been extended to support additional DFT codes, including ABACUS, making it more accessible to a broader user base. A spin Hamiltonian downfolding methodology incorporates contributions from ligand spin effects~\cite{solovyev_2021_exchange}. The spin Hamiltonian is symmetrized through the averaging of symmetry-equivalent terms. 

These enhancements in TB2J, coupled with its improved integration with the \multibinit package, provide a more robust framework for studying magnetic properties and spin dynamics in materials.

\subsection{Temperature Dependent Effective Potential}
\label{subsect:multiscale_tdep}

The TDEP method~\cite{Hellman2011} is a second-principles framework for calculating temperature-renormalized
effective lattice dynamical Hamiltonians. It serves as a post-processing tool that transforms energy and force trajectories from DFT (or other methods such as 
Machine Learning, path integral MD, or MB-EAP) into effective temperature-dependent interatomic force constants. This enables the computation of anharmonic phonons, free energies, phonon spectral functions, and lattice thermal conductivity, among other functionalities. Recently, it has been extended to incorporate dielectric response quantities~\cite{Benshalom2022}, explicitly taking temperature renormalization into consideration for infrared and Raman spectra.

TDEP was initially a post-processing tool that utilized, for instance, \abinit output to generate the effective Hamiltonians. We have introduced the capability to incorporate the temperature-renormalized phonons back into \abinit, facilitating, for example, the calculation of electron-phonon coupling in systems where the bare harmonic phonons are insufficient or unstable.

This integration is achieved through the \verb|dump_dynamical_matrices| utility within the TDEP package, which generates dynamical matrices in DDB format.



\subsection{\abinit implementation of \textsc{aTDEP}}
\label{subsect:multiscale_atdep}
A post-processing code based on the TDEP method, called \atdep, has been developed and integrated into the \abinit software under a GNU-GPL license. This tool computes various thermodynamic and vibrational properties such as phonon spectra, elastic moduli, Grüneisen parameters, thermal expansion, free energy, and heat capacity. 
It has been validated on many elemental materials, compounds and alloys,
including uranium~\cite{Bouchet2015, Bouchet2017}, 
zirconium~\cite{Anzellini2020}, \ce{MgO}~\cite{Bouchet2019},
iron~\cite{Bouchet2022}, and others, across a wide range of thermodynamic conditions, showing excellent agreement with experimental results. 
A previous paper~\cite{Bottin2020} provides theoretical background, algorithmic details, and practical examples for \atdep users. 

Implementing \atdep  in \abinit allows it 
to take advantage of existing features such as symmetry 
and \textbf{q}-point grid handling, dipole-dipole 
correction, and efficient, transferable  file formats 
(NetCDF, YAML, DDB). Among the features provided 
by the \abinit environment, two have recently benefited 
the \atdep project: the development of an online tutorial 
and the construction of a non-regression test suite, with almost 40 test inputs which are useful examples for beginning users. 

Other implementations have recently been introduced. One is the ability to parallelize calculations over equilibrium atomic configurations obtained from canonical sampling. This feature is automatically triggered when the code is executed in parallel and can also be manually activated by specifying the keyword \abivar{nproc} followed by two integers ({\em e.g.}, 1 4, to run the calculation on 4 processors). This parallelization is particularly beneficial when the number of atomic configurations reaches several thousand. Note that the computational cost can also be reduced by using the keyword \abivar{slice} followed by an integer ({\em e.g.}, 100, to select only one atomic configuration 
out of every 100) when the database originates from a molecular dynamics trajectory. In this case, the selected configurations are naturally 
uncorrelated.

Secondly, it is now possible to use statistical weights derived from canonical sampling (as in an \mlacs calculation~\cite{Castellano2022, Richard2023, Castellano2024}, see section~\ref{sec:mlacs}) by specifying the sequence ``\abivar{use\_weights} 1''. All thermodynamic quantities are then reweighted accordingly. Third, a new feature has been implemented in \atdep to handle substitutional alloys within the Virtual Crystal Approximation. This functionality can be activated by specifying the \abivar{alloy} variable in the input file (see the documentation for details). An example of this approach is provided in the study of the \ce{UMo} system~\cite{Castellano2020}. 

Furthermore, the calculation of interatomic force constants (IFCs), initially limited to second order (by default) and third order (through
the input variables \abivar{order} 3 rcut3), has been extended to include fourth-order IFCs (accessed as \abivar{order} 4 rcut3 rcut4), where rcut3 and rcut4 define the interaction cutoff radii. Different orders can be computed either simultaneously (\abivar{together} 1) or sequentially (\abivar{together} 0). 
The second-order IFCs are Fourier-transformed to yield the dynamical matrix, which is stored in a DDB file. This file can be combined with Born effective charges, and further analyzed by \abiexec{anaddb} to compute the phonon band structure and density of states.
It can also be integrated into any workflow that requires the dynamical matrix such as the electron-phonon interaction. 

Finally, the treatment of low-symmetry systems, such as rhombohedral crystals exhibiting non-symmorphic symmetries ({\em e.g.}, chalcogenides), has been completely revised and improved.


%% file: 7_highthroughput.tex
\section{High Throughput Workflows and Libraries}

\label{sect:highthroughput}



High-throughput workflows have emerged as a cornerstone in modern computational materials science over the last ten years. This effort has enabled scientists to tackle an increasingly large number of calculations with complex structures and even larger-scale simulations with improved efficiency and reproducibility. This section introduces a suite of advanced, modular, and automated workflows that streamline the entire computational process as implemented in packages that use \abinit as the core engine—from initial input generation and convergence studies to the comprehensive post-processing of simulation data.

In the discussed workflows, we demonstrate how we can bridge the gap between basic first-principles outputs and drawing meaningful scientific insights by leveraging open-source tools and frameworks such as AbiPy, Atomate2~\cite{Ganose2025}, ASE~\cite{HjorthLarsen2017}, and \aiida~\cite{Huber2020,Uhrin2021}. Our discussion addresses many challenges inherent in high-throughput calculations, as observed in other implementations, but using some of the specific features of \abinit, including managing complex input/output structures, systematic parameter studies, and integrating sophisticated analysis routines, with implemented properties. We demonstrate that within the \abipy framework, we are not only able to facilitate the automation of \abinit calculations, but additionally, we integrate  robust  libraries like Pymatgen~\cite{Ong2013} and ASE. This interface then is able to ensure that tasks such as symmetry handling, magnetic configuration generation, structural relaxation, and data visualization are executed with minimal manual intervention.

More importantly, we discuss here specific implementations—such as those for computing luminescence spectra or thermal expansion—which demonstrate how high-throughput approaches can be tailored to extract nuanced physical properties. The present modules utilize state-of-the-art techniques, like the $\Delta$SCF method for excited-state simulations or the Quasi-Harmonic Approximation (QHA) for thermal studies, alongside automated workflows that ensure consistency and reliability across multiple computational studies. The integration with protocols like i-PI for coupling \abinit with ASE further exemplifies the modularity and flexibility of these workflows, enabling advanced molecular dynamics and geometry optimization routines. In addition, incorporating platforms like Atomate2 and the \aiida plugin for \abinit not only enhances the scalability of these workflows—allowing the efficient screening of thousands of materials—but also emphasizes meticulous record-keeping and data provenance. This rigorous approach to data management is essential for reproducibility, enabling other researchers to verify and build upon existing studies.

\subsection{\abipy}
\label{subsect:highthroughput_abipy}

\abipy is an open-source Python library that facilitates the analysis, automation, and post-processing of first-principles calculations performed with \abinit. 
\abipy integrates seamlessly with pymatgen~\cite{Ong2013}, a robust materials science library, to efficiently handle crystal structures, symmetry operations, and file I/O. The package enables users to automate workflow generation, perform convergence studies, and generate publication-ready plots with minimal effort. 
Its modular architecture allows researchers to explore computational results, extract physical insights interactively, 
and optimize simulation parameters.
\abipy has been used to automate calculations or post-processing in several different studies, ranging from DFPT calculations~\cite{Petretto2018, Petretto2018a, deMelo2023},
optical properties within the Bethe-Salpeter equation~\cite{Gillet2016, Gillet2017},
$GW$ convergence studies~\cite{Setten2017}, 
transport properties within the Boltzmann formalism~\cite{Claes2022}, 
assessment of universal machine learning interatomic potentials for phonon prediction~\cite{Yu2024},
and the generation of optimized norm-conserving 
Vanderbilt pseudopotentials~\cite{Hamann2013, Setten2018, Tantardini2024}
validated against all-electron calculations~\cite{Lejaeghere2016, Bosoni2023}.


\subsubsection{Automated computation of luminescence spectra}
\label{subsect:highthroughput_PL}

A set of Python modules integrated into the \abipy framework now facilitates the automated computation of phonon-resolved luminescence spectra for defects in inorganic solids. This addresses the growing demand for efficient and reproducible calculations of defect-related luminescent properties, which are crucial for applications ranging from quantum technologies~\cite{wolfowicz2021quantum} to down-conversion phosphors used in white-light LEDs~\cite{fang2022evolutionary}. 
The workflow automates key steps in a classical computational process~\cite{li2025luminescence}, from the initial $\Delta$SCF DFT calculations with constrained occupations to simulate the excited state, to the generation of defect phonon modes in large supercells, and finally the calculation of luminescence spectra based on Huang-Rhys theory~\cite{huang1950theory}, using the generating function formalism~\cite{alkauskas2014, jin2021photoluminescence}.

\begin{figure}
	\centering
	\includegraphics[width=0.99\linewidth]{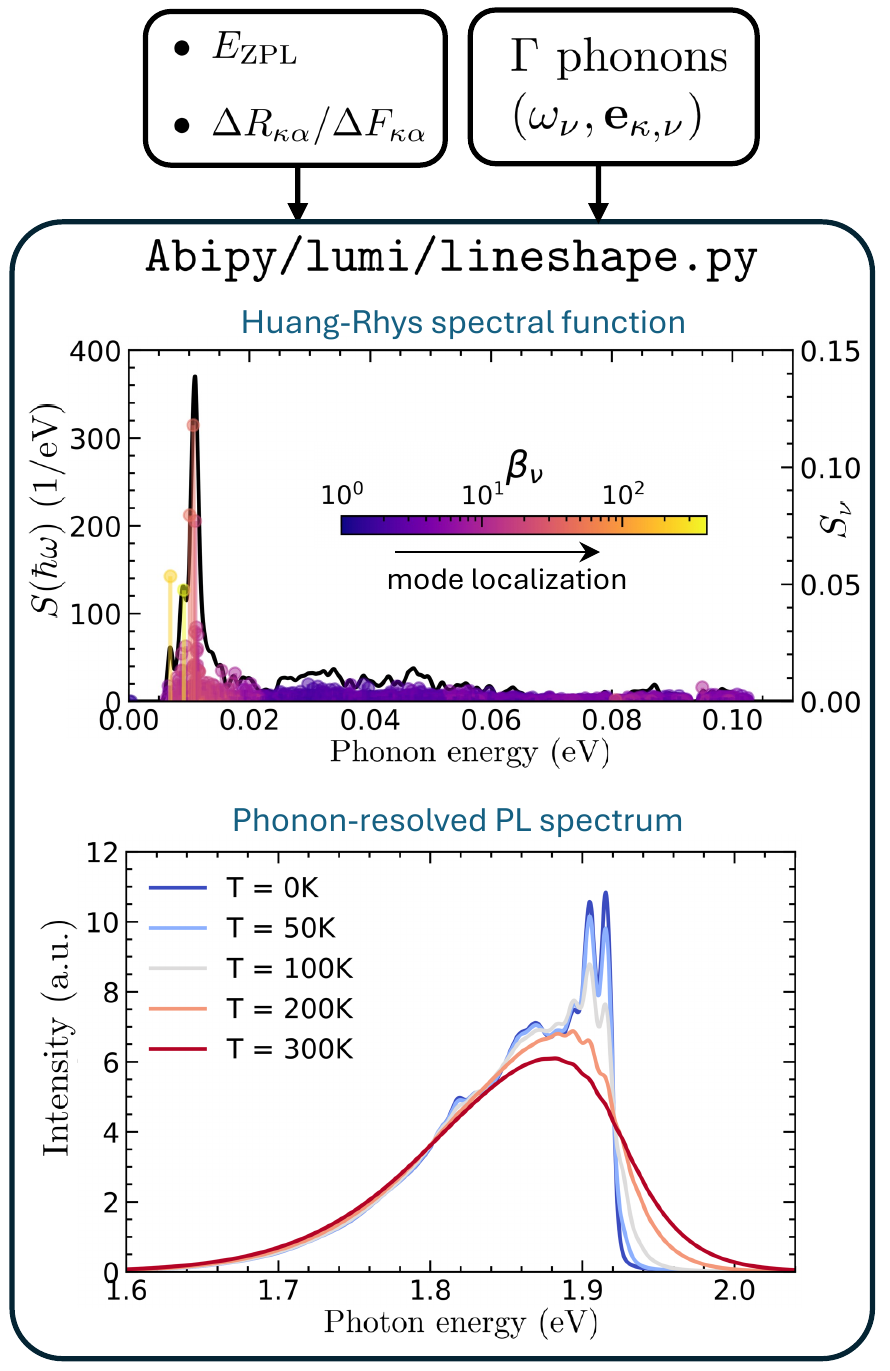}
	\caption[lineshape module]{The lineshape module enables the calculation of temperature-dependent photoluminescence spectra. Starting from the zero-phonon line energy $E_{\text{ZPL}}$, the atomic displacements $\Delta R_{\kappa\alpha}$ induced by electronic transitions, and the phonon modes, the Huang-Rhys spectral function $S(\hbar\omega)$ is computed. This function is then used to obtain the temperature-dependent PL spectra, with resolved phonon sidebands.
}
	\label{fig:lineshape_module}
\end{figure}

While these modules are designed to work together, with the output of one serving as input for the next, they can also be used independently. The process typically begins with the LumiWork module, which automates \abinit DFT tasks with $\Delta$SCF constrained occupations to compute the ground- and excited-state energies for defect systems. The $\Delta$SCF Post-Processing module then analyzes the results using a 1D configuration coordinate model to determine transition energies, Huang-Rhys factors, and effective phonon frequencies. The IFCs embedding module enables the calculation of defect phonons in large supercells by combining defect IFCs from a small supercell with pristine IFCs. Finally, the Lineshape Calculation module (see Figure~\ref{fig:lineshape_module}) integrates all components to compute the Huang-Rhys spectral function and generate temperature-dependent photoluminescence (PL) spectra. 
These developments have been applied to explain the luminescence properties of two red Eu-doped phosphor materials of technological significance~\cite{bouquiaux2021importance, bouquiaux2023first}.

\subsubsection{Thermal Expansion} 
\label{subsect:highthroughput_expansion}

Recent developments in \abipy have introduced Python modules and workflows designed to enhance the efficiency and applicability of the Quasi-Harmonic Approximation (QHA) for materials with diverse crystallographic symmetries. 
These advances focus on optimizing the computation of thermal expansion by integrating methods such as the Zero Static Internal Stress Approximation (ZSISA) and its volume-constrained variant, 
v-ZSISA~\cite{Allan1996,Taylor1997,Taylor1999,Rostami2024,Rostami2025}.
Within the QHA framework, the equilibrium structure with primitive vectors
$[R(T,P_{\textrm{ext}})]$ and associated volume
$V(T,P_{\textrm{ext}})$
at a given temperature $T$ and external pressure  $P_{\textrm{ext}}$ is 
determined by minimizing the Gibbs free energy:
\begin{multline} G([R(T,P_{\textrm{ext}})],T,P_{\textrm{ext}}) = E_{\textrm{BO}}([R(T,P_{\textrm{ext}})]) \\
    +F_{\textrm{vib}}([R(T,P_{\textrm{ext}})],T) + P_{\textrm{ext}}V(T,P_{\textrm{ext}}), 
\end{multline}
where $E_{\textrm{BO}}$ is the Born-Oppenheimer (BO) energy, $F_{\textrm{vib}}$ is the vibrational free energy computed from phonon spectra obtained by DFPT. The equilibrium structure is obtained by minimizing this Gibbs free energy with respect to all internal and external degrees of freedom at each temperature and pressure.
QHA is combined with the Zero Static Internal Stress Approximation (ZSISA) and its volume-constrained variant (v-ZSISA) to improve computational efficiency. 
In v-ZSISA, lattice vectors and atomic positions are optimized at a fixed volume for zero temperature, while in ZSISA, lattice vectors are treated as external strains, and atomic positions are internal strains. 
For each external strain, internal strains must vanish for zero temperature, significantly reducing the required phonon calculations.
Further efficiency is achieved by approximating the Taylor expansion of the vibrational free energy~\cite{Rostami2024,Rostami2025}. When Taylor expansion terms up to second order only are considered, this becomes the $E{\infty}\textrm{Vib}2$ approximation. 
This approach uses finite difference methods to compute $F_{\textrm{vib}}$ derivatives, requiring vibrational free energy calculations for specific lattice deformations. 
A dedicated \abipy module automates the generation of these deformations, followed by structural relaxation and phonon calculations, ensuring a systematic workflow.

In the post-processing step, for both v-ZSISA methods and ZSISA-$E{\infty}\textrm{Vib}2$ for uniaxial structures, the thermal expansion is determined by fitting an equation of state (EOS) or applying polynomial interpolation to the Gibbs free energy and minimizing it at different temperatures and pressures using \abipy modules. 
The ZSISA-$E{\infty}\textrm{Vib}2$ method, being more general, works across a wide range of crystallographic symmetries, from cubic to monoclinic and triclinic. However, this method requires an additional self-consistent algorithm during the post-processing step to iteratively refine the optimal structure at each temperature and pressure~\cite{Rostami2025}.

The workflow begins with an initial lattice configuration $[R]$, from which the thermal and BO stresses are computed. 
A target stress is defined based on the thermal stress and external pressure, $P_{\textrm{ext}}$. Using \abinit capabilities, the lattice parameters are optimized with a stress-target constraint. 
After each relaxation, the thermal stress is recalculated, and the target stress is adjusted accordingly. This iterative process continues until the thermal and BO stresses match, ensuring self-consistency.
Following lattice optimization, BO elastic constants are computed via DFPT~\cite{Hamann2005} to determine the thermal expansion coefficients and elastic constants at different temperatures and pressures. 
The integrated workflow in \abipy automates all of these steps, providing a systematic and efficient approach for computing temperature- and pressure-dependent structural properties.

\subsubsection{Interface between \abinit and ASE via the i-PI protocol}
\label{subsect:highthroughput_sockets}

The ASE (Atomic Simulation Environment)~\cite{HjorthLarsen2017} is an open-source Python library for establishing,
running, and analyzing atomistic simulations. 
It provides a flexible interface to various electronic structure codes, and can efficiently communicate with external codes via the protocol from the i-PI software~\cite{Ceriotti2014, Kapil2019}, 
significantly accelerating geometry optimizations, molecular dynamics, and other algorithms
where ASE controls the atomic motion. 
Advanced algorithms, including constraints specified at the Python level, can be applied with \abinit 
computing energies, forces, and stress, then returning the results to ASE.
Notably, ASE does not require the i-PI package but simply adopts its protocol.
This option can be activated by setting \abivar{ionmov} to 28.
Examples of how to use the interface between ASE and \abinit are available at 
\href{https://wiki.fysik.dtu.dk/ase/ase/calculators/socketio/socketio.html}{on this page}.

\subsection{Atomate2 interface for \abinit}
\label{subsect:highthroughput_atomate2}
Atomate2~\cite{Ganose2025} is a free, open-source software for performing complex materials science workflows using simple Python functions, built on Jobflow~\cite{Rosen2024}.
It can scale from a single material to 100 materials or 100,000 materials. It provides easy routes to modifying and chaining workflows while automatically keeping meticulous records of jobs, their directories, runtime parameters, and more.
It includes a library of "standard" workflows for many first-principles software, allowing the computation of a wide variety of desired material properties.

The \abinit interface in Atomate2 enables the execution of several tasks, including standard DFT calculations like structural relaxation (atoms and/or cells) and SCF and NSCF calculations (uniform or bandstructure). It also supports many-body perturbation theory (MBPT) calculations, specifically quasiparticle energies within the GW approximation and dielectric function calculation by solving the Bethe-Salpeter equation. Furthermore, the porting of the AbiFlows package for DFPT calculations is in progress with the phonons, optical dielectric, and static second-harmonic generation (SHG) tensor workflows. The latter already enabled a high-throughput study mixing both first-principles and machine learning screening, which resulted in a new $\sim$2200 dataset of SHG tensors at the LDA level along with $\sim$700 scissor-corrected tensors~\cite{Trinquet2025Apr}. Atomate2 improves the reproducibility of this type of work, as workflows can be serialized in a dictionary format and shared to promote reuse and verification with minimal effort.

\begin{figure}[h!]
	\centering
	\includegraphics[width=\linewidth]{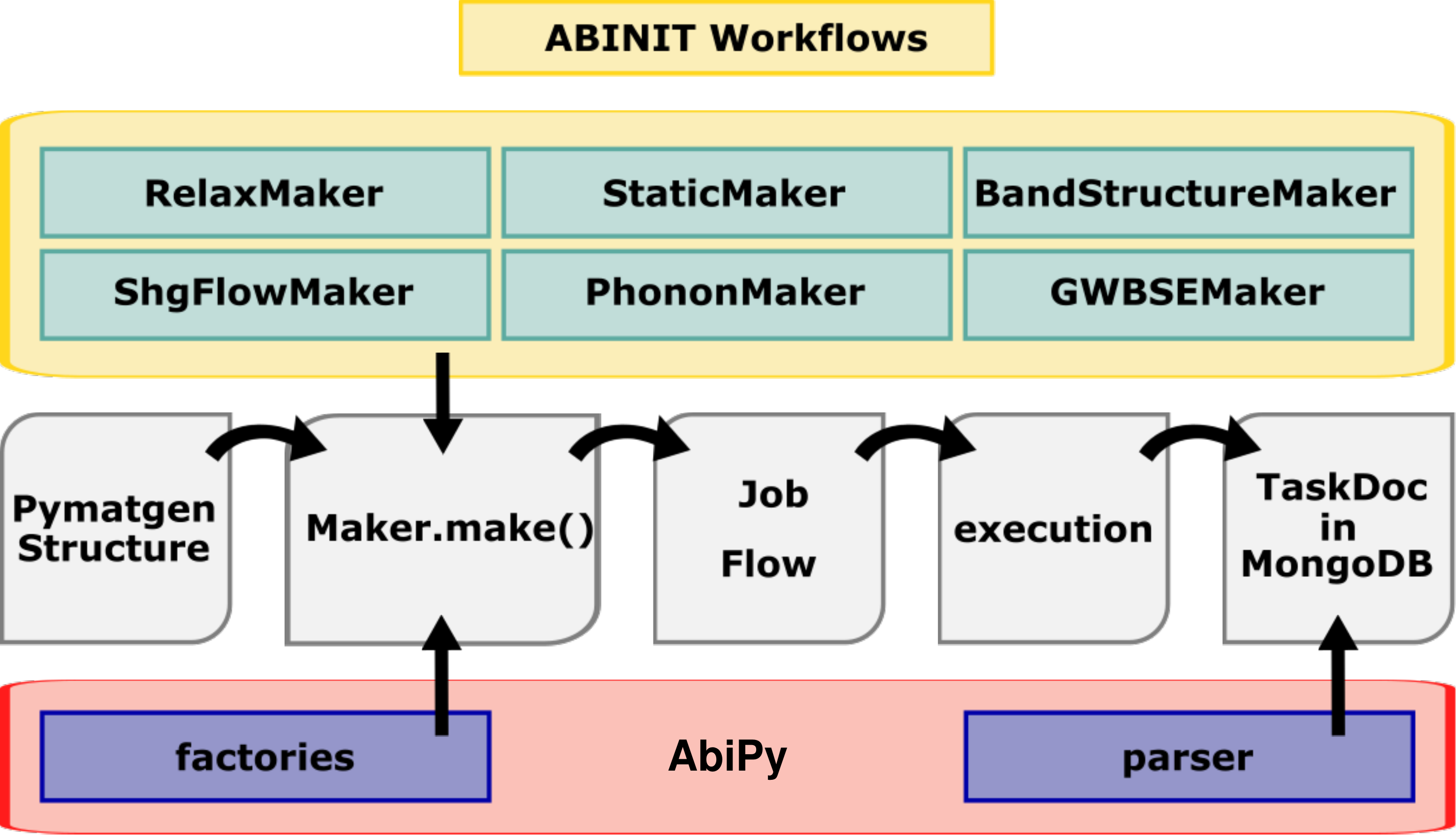}
	\caption[lineshape module]{Schematic of the implementation of \abinit workflows in atomate2. Each \texttt{Maker} (Static, Relax, BandStructure, Phonon, ShgFlow, and GWBSE) relies on \abipy factories to create the inputs and on its parser to extract the relevant information from the outputs.}
	\label{fig:atomate2}
\end{figure}

The implementation is heavily based on the functionalities of \abipy for automatic input generation and output processing. Each \texttt{Maker} or calculation type inherits from \texttt{BaseAbinitMaker} and has its own \texttt{AbinitInputGenerator}, which uses specific \abipy factory functions to create the appropriate \texttt{AbinitInput}. Upon job completion, \abipy parsing capabilities are used to retrieve relevant outputs and construct an \texttt{AbinitTaskDoc} with common fields such as \texttt{output.energy}, \texttt{output.bandgap}, and \texttt{output.forces}. It is also possible to store relevant files like DDB or NetCDF files in a FileStore partition of MongoDB for later manipulation with \abipy, such as generating plots for band structures, density of states, or spectra. Basic figures are saved automatically when feasible. By default, the \abinit workflows look for pseudopotentials from Pseudodojo in the \texttt{$\mathtt{\sim}$/.abinit/pseudos} folder, and these can be downloaded using the \abipy \abiexec{abips.py} command. Although the use of custom pseudopotentials is possible, active development is focused on improving this. The schematic representation of the Jobs execution is shown in Fig.~\ref{fig:atomate2}. 

\subsection{\aiida plugin for \abinit}
\label{subsect:highthroughput_aiida}
\aiida~\cite{Huber2020,Uhrin2021}, is an open-source high-throughput workflow framework written in Python with automated data management and full provenance tracking, the ability to saturate exascale computing resources, and broad support within the computational materials science community~\cite{Bosoni2023}, with community-developed plugins supporting most first-principles codes.
This wide range of support was recently leveraged to verify the precision of DFT implementations~\cite{Bosoni2024}, where the plugins for \abinit, BigDFT, CASTEP, CP2K, Quantum ESPRESSO, SIESTA, VASP, FLEUR, and WIEN2k have been used.
\aiida is built around a core-plugin architecture, where the workflow manager, provided by the \texttt{aiida-core} Python package, is designed to be generic with respect to the type of code being run.
Specific data types, input writers, output parsers, provenance-tracked functions, and full workflows are implemented in plugin packages.

\begin{figure*}[th]
	\centering
	\includegraphics[width=\linewidth]{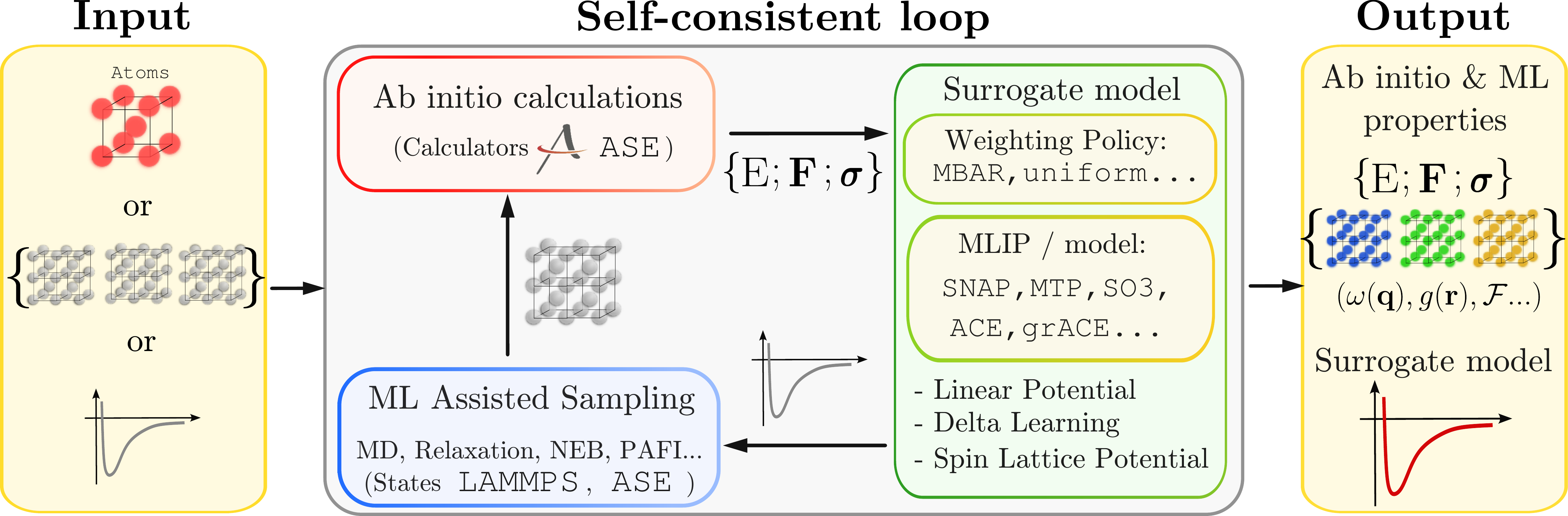}
	\caption[mlacs workflow]{Workflow of the \mlacs Code. Structure prototypes, initial interatomic potentials, and/or a set of atomic configurations are provided as input.
    In the \mlacs self-consistent loop, DFT energies (E), forces (\textbf{F}), and stresses ($\boldsymbol{\sigma}$) (from \abinit or other ASE calculators) are used to update a surrogate model. The model comprises a MLIP, 
    for instance SNAP~\cite{Thompson2015}, MTP~\cite{Novikov2021}, ACE~\cite{Drautz2019}, or GRACE~\cite{Bochkarev2024}, and a weighting scheme such as  MBAR~\cite{Shirts2008,shirts2017reweighting} or uniform. New atomic configurations are generated with the \textsc{Lammps}~\cite{Thompson2022} and/or \textsc{ASE}~\cite{ase-paper} packages. 
    The output consists of two \textit{ab initio} datasets: one containing the generated atomic configurations and the other with corresponding energies, forces, and stresses. Additional outputs include phonon frequencies, radial distribution functions, or free energies, along with the final MLIP and atomic configuration weights.}
	\label{fig:mlacs}
\end{figure*}
The \texttt{aiida-abinit} plugin (v 0.5.0) currently supports standard DFT calculations, including structural relaxation, SCF, and NSCF calculations.
However, unlike the Atomate2 interface, it does not yet support MBPT, GW, or other advanced calculations.
Parsers for the \texttt{abo} main output file, GSR, and HIST outputs are implemented; additional output parsing supported by \abipy can be easily added when needed.
%
One key feature of \aiida is the ability to automate error handling and \texttt{aiida-abinit}  currently reports some common simple error states, including missing files or folders, out-of-walltime, maximum number of scf or ionic minimization cycles reached, and the presence of other errors and warnings in the output file.
Out-of-wall-time errors are handled via an automatic restart from output for SCF calculations or by launching a new calculation with the most recent atomic structure for relax calculations.
Some errors, such as missing files, are considered unrecoverable and require user intervention.
The \texttt{aiida-abinit} plugin depends on \abipy for core I/O functions, including writing \abinit input files and parsing output files.
Therefore, any valid input which can be defined by \abipy \texttt{AbinitInput} class is supported.
However, to achieve better interface consistency, atomic structures, $\kk$-points, and pseudopotentials are managed via \aiida before being provided to \abipy.
The atomic structure is provided using \aiida's \texttt{StructureData} class, which provides functions for conversion from both ASE \texttt{Atoms} and Pymatgen \texttt{Structure} and \texttt{Molecule} classes.
K-points are managed via \aiida's \texttt{KpointsData} class, which supports regular and shifted Monkhorst-Pack meshes and explicit lists of $\kk$-points.
Pseudopotentials are handled through the \texttt{aiida-pseudo} plugin, currently in version 1.7.0, which supports the PseudoDojo~\cite{Setten2018} and SSSP~\cite{Prandini2018} tables with automated recipes and user-provided pseudopotentials through manual import.
By default, \texttt{aiida-abinit} will use the suggested kinetic energy cut-offs provided by the automatically-installed PseudoDojo and SSSP pseudopotential families; users can always override these selections and define their own cut-offs.

In its current state, \texttt{aiida-abinit} is sufficient for performing moderately-sized (1,000s) high-throughput studies requiring simple DFT calculations, but is limited by the number of output files supported for parsing and the minimal amount of error handling implemented.
The next steps to enable more user-friendly and robust calculations are to implement input protocols, expand output parsing support, and expand error handlers.
Input protocols are a concept introduced in \aiida by the Quantum ESPRESSO plugin, which allows for predefined input parameter recipes that target different precision and accuracy goals.
Robust and well-tested protocols reduce the burden on the end user and help to ensure the reliability of simulation results generated by the plugin.
Expanding the number of output files supported for parsing in \aiida will naturally allow for more advanced use cases of the plugin and will be necessary to expand its abilities beyond simple single-point and structure relaxation calculations.
Finally, the number of common errors that are properly handled by the plugin should be extended in order to ensure the plugin's ability to run large-scale high-throughput calculations. 

\subsection{Machine learning assisted canonical sampling (\textsc{MLACS}) \label{sec:mlacs}}

\mlacs~\cite{Richard2023, Castellano2024} is an open-source Python package designed to implement the Machine-Learning Assisted Canonical Sampling (\mlacs) method~\cite{Castellano2022} based on an active learning strategy.
Its primary feature is the ability to sample the canonical ensemble associated with DFT potential energy surfaces and generate equilibrium atomic configurations.
This is achieved through a self-consistent variational procedure that employs a surrogate distribution constructed from machine-learning interatomic potentials (MLIPs), following the workflow presented in Fig.~\ref{fig:mlacs}.
This approach enables the computation of finite temperature properties at a significantly reduced cost compared to {\it ab initio} molecular dynamics~\cite{Bottin2024, Siberchicot2025}.
In addition to its core functionality, \mlacs accelerates various other simulations by using MLIPs as surrogate potentials.
These simulations include energy minimization, nudged elastic band calculations, or thermodynamic integration.
The package supports several MLIP models, including SNAP~\cite{Thompson2015}, MTP~\cite{Novikov2021}, ACE~\cite{Drautz2019}, or GRACE~\cite{Bochkarev2024}, and is designed to automate the various steps required for these surrogate-based simulations.
It provides a seamless interface between the different codes necessary to perform these tasks.

The framework is compatible with any DFT code that interfaces with ASE.
However, special attention has been paid to ensure robust integration with the \abinit ecosystem.
For example, \mlacs includes a dedicated interface to run \abinit, handling both the writing and reading of the input and output files.
The package is also interfaced with \abinit's NetCDF output files, enabling the reading of OUT, GSR and HIST files after a single DFT computation. In addition, \mlacs can also read HIST files from molecular dynamics simulations and convert them directly into ASE trajectories, making it simple to train a MLIP from existing \abinit trajectories, for example. Moreover, \mlacs generates its own HIST file on-the-fly, allowing the use of tools such as \textsc{qAgate} during \mlacs execution and ensuring compatibility with \textsc{aTDEP}.

%% file: 8_conclusion.tex
\section{Conclusions}
\label{sect:conclusion}

The updates and enhancements introduced to the \abinit software since 2020 have substantially expanded its capabilities, reinforcing its position as a widely adopted tool in computational materials science. 
Since its initial release in 1998, \abinit has continuously evolved, integrating new features and performance improvements and pushing advances in the field.
This sustained growth emphasizes the software's flexible architecture and the ongoing dedication of its developers to supporting researchers worldwide.

The highlights presented in this paper span multiple domains. Foremost among these are enhancements in ground-state calculations, for example through Constrained Density Functional Theory (CDFT) using Lagrange multiplier methods. It is possible to pin down charges and magnetic moments exactly, and confidently probe excited states and local excitations. 
\abinit has also pushed DFT beyond room temperature, adding thermal exchange–correlation functionals and an extended plane-wave treatment to model warm dense matter and plasma conditions. 
\abinit DFPT routines now output phonon angular momentum and include a unique implementation of spatial dispersion with clamped‐ion and lattice‐mediated flexoelectric coefficients. 
Together, these features allow first‐principles prediction of natural optical rotation and strain‐gradient polarization capabilities which are particularly useful for low‐symmetry and chiral materials.
Excited state formalisms beyond DFT have long been a strong point of \abinit and now include a cutting-edge large-scaling GW module, Coupled Cluster calculations through the Cc4s code, and extensive options in Dynamical Mean Field Theory. 

\abinit core kernels have been extended, including SCF loops, FFTs and dense linear-algebra routines, to run on GPUs via OpenMP offload and calls to vendor libraries (cuFFT/cuBLAS on NVIDIA, rocFFT/rocBLAS on AMD) for the most costly transforms and matrix multiplications. Compared to a CPU run, the computational time is typically reduced by one order of magnitude on modern A100 or MI100 cards, for a system of about one thousand atoms using the same number of nodes.
The use of GPUs makes it possible to converge self-consistent ground states on systems of hundreds or thousands atoms in wall-clock times that were previously out of reach.

In the coming years \abinit will push further developments of advanced methodologies in particular extending support for relativistic and non-collinear states in heavy and/or magnetic materials. The GPU acceleration will be broadened to other run-modes, and adapted to new architectures and evolving HPC standards. The RT-TDDFT will implement non-adiabatic MD and responses to laser pulses. A GW perturbation theory (GWPT) implementation will be completed, a first in a single integrated code.

The core values of the project and team are the open source philosophy, promoting scientific research throughout the globe, fostering an inviting environment for new developers and contributors, and collaborations with other initiatives, packages, libraries and networks in electronic structure and materials science.
The ongoing development of ABINIT is guided by the goal of providing a robust and versatile platform for electronic structure research. 
Our efforts will continue to focus on making the code reliable, accessible, and well-documented, with increased capabilities spanning from standard DFT calculations to advanced many-body techniques, including GW and DMFT. As electronic structure theory advances, \abinit will evolve to incorporate new methods and expand its applications.

%% file: acknowledgements.tex
\begin{acknowledgments}
This work has been supported by many funding agencies and grants.
We acknowledge support from the Fonds de la Recherche Scientifique (FRS-FNRS Belgium) through the PdR Grant No. T.0103.19 - ALPS .
This work is an outcome of the Shapeable
2D magnetoelectronics by design project (SHAPEme, EOS
Project No. 560400077525) that has received funding from
the FWO and FRS-FNRS under the Belgian Excellence of
Science (EOS) program.
This work was
supported by the Communaut\'e fran\c{c}aise de Belgique through
the SURFASCOPE project (ARC 19/24-102).
S. P. is a Research Associate of the Fonds de la Recherche Scientifique - FNRS.
This work was supported by the Fonds de la Recherche Scientifique - FNRS under Grants number T.0183.23 (PDR) and  T.W011.23 (PDR-WEAVE). 
This publication was supported by the Walloon Region in the strategic axe FRFS-WEL-T.
The Flatiron Institute is a division of the Simons Foundation.

LM and DDO'R acknowledge the support of Trinity College Dublin through its Provost’s PhD Project Awards and School of Physics.

F.G.O. acknowledges financial support from MSCA-PF 101148906 funded by the European Union and the Fonds de la Recherche Scientifique (FNRS) through the grant FNRS-CR 1.B.227.25F.

This work was supported by the European Union’s Horizon 2020 research and innovation program under grant agreement number 964931 (TSAR) and by F.R.S.-FNRS Belgium under PDR grants T.0107.20 (PROMOSPAN) and T.0128.25 (TOPOTEX).

This work was supported by the European Union’s
Horizon 2020 research and innovation program under
the grant agreement N° 951786 (NOMAD CoE).

The authors acknowledge extensive access to computing time for testing and scaling on the following platforms.
The Consortium d'Equipements de Calcul Intensif (CECI funded by FRS-FNRS Belgium Grant No. 2.5020.11);
the Lucia Tier-1 of the F\'ed\'eration Wallonie-Bruxelles (Walloon Region grant agreement No. 1117545).
\end{acknowledgments}